\documentclass[prd,
               twocolumn,
               eqsecnum,
               showpacs,
               letterpaper,
               superscriptaddress,
               altaffilletter]
               {revtex4}

\usepackage{color}
\usepackage{graphicx}
\usepackage{latexsym}
\usepackage{float}
\usepackage{amsmath}

\renewcommand{\today}{\number\day\space\ifcase\month\or
  January\or February\or March\or April\or May\or June\or
  July\or August\or September\or October\or November\or December\fi
  \space\number\year}

\def\be{\begin{equation}}
\def\ee{\end{equation}}
\def\bi{\begin{itemize}} 
\def\ei{\end{itemize}}
\def\ben{\begin{enumerate}}
\def\een{\end{enumerate}}
\def\hrssu{Hz${}^{-1/2}$}
\def\hrss{h_\mathrm{rss}}

\newcommand\ligodoc{P080056-v12}

\begin{document}

\title{Search for gravitational-wave bursts in the first year of the fifth LIGO science run}

\begin{abstract}  
\vspace*{0.2in}
We present the results obtained from an  all-sky search for
gravitational-wave (GW) bursts in the 64--2000\,Hz frequency range in data
collected by the LIGO detectors during the first year (November 2005 --
November 2006) of their fifth science run.The total analyzed livetime was 268.6 days.
Multiple hierarchical data analysis methods were invoked in this search.
The overall sensitivity expressed in terms of the root-sum-square
(rss) strain amplitude $\hrss$ for gravitational-wave bursts with various
morphologies was in the range of $6 \times 10^{-22}$ \hrssu\ to a few $ \times 10^{-21}$ \hrssu.
No GW signals were observed and a frequentist upper limit 
of 3.75 events per year on the rate of  strong GW bursts was placed at the 90\% confidence level. 
As in our previous searches, we also combined this rate limit with the
detection efficiency for selected waveform morphologies 
to obtain event rate versus strength exclusion curves.
In sensitivity, these exclusion curves are the most stringent to date. 

\end{abstract}

\pacs{
04.80.Nn, 
07.05.Kf, 
95.30.Sf, 
95.85.Sz  
}

\newcommand*{\AG}{Albert-Einstein-Institut, Max-Planck-Institut f\"{u}r Gravitationsphysik, D-14476 Golm, Germany}
\affiliation{\AG}
\newcommand*{\AH}{Albert-Einstein-Institut, Max-Planck-Institut f\"{u}r Gravitationsphysik, D-30167 Hannover, Germany}
\affiliation{\AH}
\newcommand*{\AU}{Andrews University, Berrien Springs, MI 49104 USA}
\affiliation{\AU}
\newcommand*{\AN}{Australian National University, Canberra, 0200, Australia}
\affiliation{\AN}
\newcommand*{\CH}{California Institute of Technology, Pasadena, CA  91125, USA}
\affiliation{\CH}
\newcommand*{\CA}{Caltech-CaRT, Pasadena, CA  91125, USA}
\affiliation{\CA}
\newcommand*{\CU}{Cardiff University, Cardiff, CF24 3AA, United Kingdom}
\affiliation{\CU}
\newcommand*{\CL}{Carleton College, Northfield, MN  55057, USA}
\affiliation{\CL}
\newcommand*{\CS}{Charles Sturt University, Wagga Wagga, NSW 2678, Australia}
\affiliation{\CS}
\newcommand*{\CO}{Columbia University, New York, NY  10027, USA}
\affiliation{\CO}
\newcommand*{\ER}{Embry-Riddle Aeronautical University, Prescott, AZ   86301 USA}
\affiliation{\ER}
\newcommand*{\EU}{E\"{o}tv\"{o}s University, ELTE 1053 Budapest, Hungary}
\affiliation{\EU}
\newcommand*{\HC}{Hobart and William Smith Colleges, Geneva, NY  14456, USA}
\affiliation{\HC}
\newcommand*{\IA}{Institute of Applied Physics, Nizhny Novgorod, 603950, Russia}
\affiliation{\IA}
\newcommand*{\IU}{Inter-University Centre for Astronomy  and Astrophysics, Pune - 411007, India}
\affiliation{\IU}
\newcommand*{\HU}{Leibniz Universit\"{a}t Hannover, D-30167 Hannover, Germany}
\affiliation{\HU}
\newcommand*{\CT}{LIGO - California Institute of Technology, Pasadena, CA  91125, USA}
\affiliation{\CT}
\newcommand*{\LO}{LIGO - Hanford Observatory, Richland, WA  99352, USA}
\affiliation{\LO}
\newcommand*{\LV}{LIGO - Livingston Observatory, Livingston, LA  70754, USA}
\affiliation{\LV}
\newcommand*{\LM}{LIGO - Massachusetts Institute of Technology, Cambridge, MA 02139, USA}
\affiliation{\LM}
\newcommand*{\LU}{Louisiana State University, Baton Rouge, LA  70803, USA}
\affiliation{\LU}
\newcommand*{\LE}{Louisiana Tech University, Ruston, LA  71272, USA}
\affiliation{\LE}
\newcommand*{\LL}{Loyola University, New Orleans, LA 70118, USA}
\affiliation{\LL}
\newcommand*{\MT}{Montana State University, Bozeman, MT 59717, USA}
\affiliation{\MT}
\newcommand*{\MS}{Moscow State University, Moscow, 119992, Russia}
\affiliation{\MS}
\newcommand*{\ND}{NASA/Goddard Space Flight Center, Greenbelt, MD  20771, USA}
\affiliation{\ND}
\newcommand*{\NA}{National Astronomical Observatory of Japan, Tokyo  181-8588, Japan}
\affiliation{\NA}
\newcommand*{\NO}{Northwestern University, Evanston, IL  60208, USA}
\affiliation{\NO}
\newcommand*{\RI}{Rochester Institute of Technology, Rochester, NY  14623, USA}
\affiliation{\RI}
\newcommand*{\RA}{Rutherford Appleton Laboratory, HSIC, Chilton, Didcot, Oxon OX11 0QX United Kingdom}
\affiliation{\RA}
\newcommand*{\SJ}{San Jose State University, San Jose, CA 95192, USA}
\affiliation{\SJ}
\newcommand*{\SM}{Sonoma State University, Rohnert Park, CA 94928, USA}
\affiliation{\SM}
\newcommand*{\SE}{Southeastern Louisiana University, Hammond, LA  70402, USA}
\affiliation{\SE}
\newcommand*{\SO}{Southern University and A\&M College, Baton Rouge, LA  70813, USA}
\affiliation{\SO}
\newcommand*{\SA}{Stanford University, Stanford, CA  94305, USA}
\affiliation{\SA}
\newcommand*{\SR}{Syracuse University, Syracuse, NY  13244, USA}
\affiliation{\SR}
\newcommand*{\PU}{The Pennsylvania State University, University Park, PA  16802, USA}
\affiliation{\PU}
\newcommand*{\UM}{The University of Melbourne, Parkville VIC 3010, Australia}
\affiliation{\UM}
\newcommand*{\MI}{The University of Mississippi, University, MS 38677, USA}
\affiliation{\MI}
\newcommand*{\SF}{The University of Sheffield, Sheffield S10 2TN, United Kingdom}
\affiliation{\SF}
\newcommand*{\TA}{The University of Texas at Austin, Austin, TX 78712, USA}
\affiliation{\TA}
\newcommand*{\TC}{The University of Texas at Brownsville and Texas Southmost College, Brownsville, TX  78520, USA}
\affiliation{\TC}
\newcommand*{\TR}{Trinity University, San Antonio, TX  78212, USA}
\affiliation{\TR}
\newcommand*{\BB}{Universitat de les Illes Balears, E-07122 Palma de Mallorca, Spain}
\affiliation{\BB}
\newcommand*{\UA}{University of Adelaide, Adelaide, SA 5005, Australia}
\affiliation{\UA}
\newcommand*{\BR}{University of Birmingham, Birmingham, B15 2TT, United Kingdom}
\affiliation{\BR}
\newcommand*{\FA}{University of Florida, Gainesville, FL  32611, USA}
\affiliation{\FA}
\newcommand*{\GU}{University of Glasgow, Glasgow, G12 8QQ, United Kingdom}
\affiliation{\GU}
\newcommand*{\MD}{University of Maryland, College Park, MD 20742 USA}
\affiliation{\MD}
\newcommand*{\AM}{University of Massachusetts - Amherst, Amherst, MA 01003, USA}
\affiliation{\AM}
\newcommand*{\MU}{University of Michigan, Ann Arbor, MI  48109, USA}
\affiliation{\MU}
\newcommand*{\MN}{University of Minnesota, Minneapolis, MN 55455, USA}
\affiliation{\MN}
\newcommand*{\OU}{University of Oregon, Eugene, OR  97403, USA}
\affiliation{\OU}
\newcommand*{\RO}{University of Rochester, Rochester, NY  14627, USA}
\affiliation{\RO}
\newcommand*{\SL}{University of Salerno, 84084 Fisciano (Salerno), Italy}
\affiliation{\SL}
\newcommand*{\SN}{University of Sannio at Benevento, I-82100 Benevento, Italy}
\affiliation{\SN}
\newcommand*{\SH}{University of Southampton, Southampton, SO17 1BJ, United Kingdom}
\affiliation{\SH}
\newcommand*{\SC}{University of Strathclyde, Glasgow, G1 1XQ, United Kingdom}
\affiliation{\SC}
\newcommand*{\WA}{University of Western Australia, Crawley, WA 6009, Australia}
\affiliation{\WA}
\newcommand*{\UW}{University of Wisconsin-Milwaukee, Milwaukee, WI  53201, USA}
\affiliation{\UW}
\newcommand*{\WU}{Washington State University, Pullman, WA 99164, USA}
\affiliation{\WU}

\author{}    \affiliation{\GU}    
\author{B.~P.~Abbott}    \affiliation{\CT}    
\author{R.~Abbott}    \affiliation{\CT}    
\author{R.~Adhikari}    \affiliation{\CT}    
\author{P.~Ajith}    \affiliation{\AH}    
\author{B.~Allen}    \affiliation{\AH}  \affiliation{\UW}  
\author{G.~Allen}    \affiliation{\SA}    
\author{R.~S.~Amin}    \affiliation{\LU}    
\author{S.~B.~Anderson}    \affiliation{\CT}    
\author{W.~G.~Anderson}    \affiliation{\UW}    
% ...
 \collaboration{The LIGO Scientific Collaboration, http://www.ligo.org}
 \noaffiliation
\date[\relax]{Dated: \today }

\maketitle

\section{Introduction}\label{sec:introduction}

After many years of preparation, interferometric gravitational wave (GW)
detectors have now begun an era of long-duration observing. 
The three detectors of the Laser Interferometer
Gravitational-Wave Observatory (LIGO)~\cite{S5} reached
their design sensitivity levels in 2005 and began a ``science run''
that collected data through late 2007.  This run is called ``S5''
since it followed a sequence of four shorter science runs that began in
2002.  The German/British GEO600 detector
~\cite{grote2008} joined
the S5 run in January 2006, and the Italian/French Virgo
detector~\cite{acernese2006} began its first science run (denoted
VSR1) in May 2007, overlapping the last $4.5$ months of the S5 run.
The data collected by these detectors provide the best opportunity
yet to identify a GW signal---though detection is still far
from certain---and is a baseline for future coordinated data
collection with upgraded detectors.

Gravitational waves in the frequency band of LIGO and the other
ground-based detectors may be produced by a variety of astrophysical
processes~\cite{CutlerThorne}. 
See for example~\cite{blanchet06} for inspiralling compact
binaries,~\cite{jaranowski1998} for spinning neutron stars,~\cite{baker2006}
for binary mergers,
and~\cite{ott04,ott06,ott2007,ott2008} for core-collapse supernovae.  

The GW waveform emitted by a compact binary system during the inspiral phase can
be calculated accurately in many cases, allowing searches with optimal
matched filtering; see, for example, \cite{S3S4inspiral}.  The
waveform from the subsequent merger of two black holes
is being modeled with ever-increasing
success using numerical relativity calculations, but is highly
dependent on physical parameters and the properties of strong-field
gravity. The uncertainties for the waveforms of other transient sources 
 are even larger. It is thus desirable to explore more generic search
algorithms capable of detecting a wide
range of short-duration GW signals from poorly-modeled sources---such as stellar core
collapse to a neutron star or black hole---or unanticipated sources.
As GW detectors extend the sensitivity frontier, it is important to
not rely too heavily on assumptions about source astrophysics or about
the true nature of strong-field gravity, and to search as broadly as
possible.

In this paper, we report on a search for GW ``burst'' signals in the
LIGO data that were collected during the first 12 months of the S5
science run. A search for GW bursts in the remainder of
the S5 data set, along with the Virgo VSR1 data, will be published
jointly by the LSC and Virgo collaborations at a later date.
 
The GW burst signals targeted are assumed to have signal
power within LIGO's frequency band and durations shorter than
$\sim$1\,s, but are otherwise arbitrary. 
This analysis, like most of our previously published searches for
GW bursts, focuses on low frequencies---in
this case 64\,Hz to 2000\,Hz---where the detectors are the most
sensitive.
A dedicated search for bursts above 2000\,Hz is presented in a companion paper \cite{S5burstAllSkyH}.

Interferometric GW detectors collect stable,
high-sensitivity (``science mode'') data typically for several hours
at a time, with interruptions due to adverse environmental conditions,
maintenance, diagnostics, and the need to occasionally regain the
``locked'' state of the servo controls.
In this analysis we searched the data at all times when
two or more LIGO detectors were operating, a departure from the
all-sky GW burst searches from earlier science
runs~\cite{S2burstAllSky,S2burstLigoTama,S3burstAllSky,S4burstAllSky,S4LIGOGEO}, which
required coincidence among three (or more) detectors.
In this paper, the term
``network'' is used to describe a set of detectors operating in science
mode at a given time. A network may include any combination of the Hanford
4\,km (H1) and 2\,km (H2) detectors,  the Livingston 4\,km (L1) detector and GEO600.
Because the GEO600 detector was significantly less sensitive
than LIGO during the S5 run
(a factor of 3 at 1000\,Hz, and almost two orders of magnitude at 100\,Hz),
we do not use its data in the initial search but reserve it
for evaluating any event candidates found in the LIGO data.

 This paper presents results from three different
``analysis pipelines'', each representing a complete search.  While
the pipelines analyzed the data independently, they began with a common 
selection of good-quality data
and applied a common set of vetoes to reject identifiable artifacts.
Each pipeline was tuned to 
maximize the sensitivity to simulated GW signals
while maintaining a fixed, low false alarm rate.
The tuning of the pipelines, the choice of good data and
the decision on the veto procedure were made
before looking at potential candidates.
 
No GW signal candidates were identified by any of the analysis pipelines with
the chosen thresholds.  In order to interpret this non-detection, we
evaluate the sensitivity of each pipeline for simulated signals of
various morphologies, randomly distributed over the sky and over time.
As expected, there are some sensitivity differences among the
pipelines, although the sensitivities rarely differ by more than a
factor of 2 (see section~\ref{sec:simulations})
and no single pipeline performs best for all of the
simulated signals considered.  We combine the results of the pipelines 
to calculate
upper limits on the rate of GW bursts as a function of signal
morphology and strength.

The rest of the paper is organized as follows:
After specifying the periods of data, forming the first year of the S5 science run in Sec.~\ref{sec:ligos5run},
Sec.~\ref{sec:s5detectors} describes the state of the detectors during that period.
Section IV summarizes the elements of this GW burst search which are common to all of the analysis pipelines.
The analysis pipelines themselves are detailed in Sec.~\ref{sec:algorithms} and Appendices C, D and~E.
Section~\ref{sec:tuning} describes how each pipeline is tuned, while Sec.~\ref{sec:simulations}
presents the sensitivity curves for simulated signals and Sec.~\ref{sec:systematics} describes the systematic errors in these sensitivity curves.
The results of the search are given in Sec.~\ref{sec:results},
and some discussion including estimates of the astrophysical reach for burst candidates 
in Sec.~\ref{sec:summaries}.

\section{S5 First-Year Data Set}\label{sec:ligos5run}

The search described in this paper uses data from approximately
the first calendar
year of S5, specifically from November 4, 2005 at 16:00 UTC
through November 14, 2006 at 18:00 UTC.

\begin{figure}
\includegraphics[width=2in]{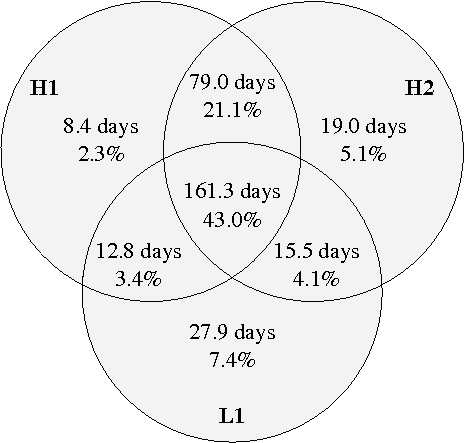}\\
\includegraphics[width=2in]{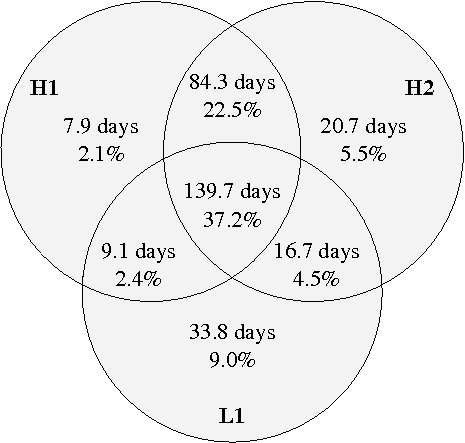}
\caption{
The top diagram indicates the mutually exclusive livetimes and duty cycles 
of different networks available for detection searches.
The category 1 and 2 data quality flags (DQF) and vetoes described in 
Appendices A and B have been applied.
The bottom diagram indicates the mutually exclusive livetimes and duty cycles 
of the different networks after category 3 DQF and vetoes have been
applied to define the data set used to calculate upper limits.}
\label{figure:venn}
\end{figure}

Figure~\ref{figure:venn} shows the amount of science-mode data collected
(``livetime'') for each mutually-exclusive network of detectors along
with percentages of the experiment calendar duration (duty cycle). The top Venn diagram represents the
data with basic data quality and veto conditions (see Sec. IV and
Appendices~\ref{sec:dataquality} and~\ref{sec:vetoes}), including
$268.6$ days of data during which two or more LIGO detectors were in
science mode; this is the sample which is searched for GW burst
signals.  An explicit list of the analyzed intervals after category 2 DQFs is 
available at~\cite{hugheyDQF}.
The bottom Venn diagram shows the livetimes after the
application of additional data quality cuts and vetoes that provide
somewhat cleaner data for establishing upper limits on GW burst event
rates.  In practice, only the H1H2L1 and H1H2 (not L1)
networks---encompassing most of the livetime, 224 days---are used to
set upper limits.

\section{The detectors}\label{sec:s5detectors}

\subsection{LIGO}

\begin{figure}
\includegraphics[width=3.25in]{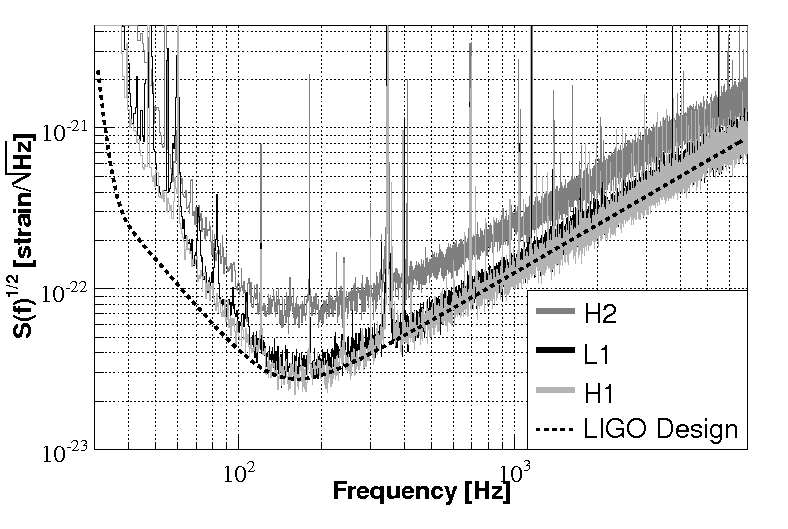}
\caption{Representative sensitivities of the LIGO detectors during the first
year of S5.  These curves show the amplitude spectral density of LIGO noise
converted to GW strain units.
}
\label{fg:ligo-noise}
\end{figure}

The high sensitivity (see Fig.~\ref{fg:ligo-noise}) and duty cycles (78.0\% for H1, 78.5\% for H2, and 66.9\% for L1) achieved during the S5 run were the
result of a number of improvements made prior to the
run~\cite{sigg2006, waldman2006}.  The major changes were the successful operation 
at Livingston of a hydraulic external pre-isolator (HEPI) to
suppress seismic disturbances, and the implementation at both sites of a
thermal compensation system (TCS) to reduce thermal lensing effects in the
interferometer arm cavities due to optical absorption in mirror coatings and substrates.
The HEPI system provides a reduction of the seismic noise by an order of
magnitude in the band 0.2--2.0\,Hz, and thus significantly improves the duty
cycle of the L1 detector.

Another significant improvement was the extension of the 
wave-front sensing (WFS) subsystem to control all
alignment degrees of freedom
of the core interferometer optics, leading to significantly reduced
alignment fluctations.  Several
improvements were made to the length sensing and control subsystem,
enabling the photodetectors to take more
power without saturation and thus allowing the laser power to be
increased.  A new method to calibrate the detectors was
introduced, based on direct actuation of the test masses via
radiation pressure from an auxiliary laser beam. Unlike the traditional
coil-drive calibration method \cite{Adhikari2003}, which requires rather
large test mass displacements, the new technique allows calibration of the
detectors at a level closer to the anticipated signal strength. 

Other improvements included modifications to acoustic and seismic isolation of
optical tables with detection photodiodes, changes to the safety shutters to
protect photodiodes from damage when interferometers fall out of lock, and improved
detection of impending saturation of photodiodes to prevent lock losses.
Finally, a
number of physical effects which led to spurious transients and spectral lines
in the data during previous science runs have been diagnosed and mitigated.

\subsection{GEO600}

The GEO600 detector, located near Hannover, Germany, was also operational during the
S5 run, though with a lower sensitivity than the LIGO detectors. The GEO600 data were not
used in the current study as the modest gains in the sensitivity to GW signals
would not have offset the increased complexity of the analysis. The GEO600 data
were held in reserve, and could have been used to follow up on detection candidates from
the LIGO-only analysis.

GEO600 began its participation in S5 on January~21,
2006, operating in a night-and-weekend mode. In this mode,
science data were acquired during nights and weekends while
commissioning work was performed during the day time. The
commissioning work focused mainly on gaining a better
understanding of the detector and improving data quality.
It was performed in a manner that avoided disrupting science 
periods and allowed for well-calibrated data to be acquired.
Between May~1 and October~6, 2006, GEO600 operated in
so-called 24/7-mode, 
during which the detector's duty cycle in science-mode operation
was maximized and only very short maintenance periods took place.
Overall in 24/7-mode an instrumental duty cycle
of about 95\% and a science-mode duty cycle of more than 90\% were
achieved.
GEO600 returned to
night-and-weekend mode on October~16, 2006, and work
began on further improving the reliability of the instrumentation
and reducing the glitch rate.
The detector was operated in night-and-weekend mode until the end
of S5 in October 2007.
Overall, GEO600 collected about 415 days of well-calibrated and
characterized science data
in the period between January 2006 and October 2007.
\section{Analysis pipeline overview}\label{sec:procedure}

In this search for GW bursts, three independent end-to-end analysis
pipelines have been used to analyze the data.  These pipelines were
developed and implemented separately, building upon many of the
techniques that were used
in previous searches for bursts in the S1, S2, S3 and S4
runs of LIGO and
GEO600~\cite{S1burstAllSky,S2burstAllSky,S3burstAllSky,S4burstAllSky,S4LIGOGEO},
and prove to have comparable sensitivities (within a
factor of $\sim$$2$; see Sec.~\ref{sec:simulations}).
One of these pipelines is fully coherent in the sense
of combining data (amplitude and phase) from all detectors and accounting
appropriately for time delays and antenna responses for a hypothetical
gravitational-wave burst impinging upon the network.
This provides a powerful test to distinguish GW signals
from noise fluctuations.

Here we give an overview of
the basic building blocks common to all of the pipelines.
The detailed operation of each pipeline will be described later.

\subsection{Data quality evaluation}

Gravitational-wave burst searches are occasionally affected by
instrumental or data acquisition problems as well as periods of
degraded sensitivity or nonstationary noise due to bad weather or
other environmental conditions.  These may produce
transient signals in the data and/or may complicate the evaluation of
the significance of other candidate events.  Conditions which may
adversely affect the quality of the data are catalogued during and
after the run by defining ``data quality flags'' (DQFs) for lists of
time intervals.  DQFs are categorized according to their
seriousness; some are used immediately to select the data to be
processed by the analysis pipelines (a subset of the nominal
science-mode data), while others are applied during post-processing.
These categories are described in more detail in Appendix A.
In all cases the DQFs were defined and categorized before analyzing
unshifted data to identify event candidates.

\subsection{Search algorithms}

Data that satisfies the initial selection
criteria are passed to algorithms that perform the signal-processing
part of the search, described in the following section and in three
appendices. These algorithms decompose the data stream
into a time-frequency representation and look for statistically significant
transients, or ``triggers''.  Triggers are accepted over a frequency
band that spans from 64\,Hz to 2000\,Hz. The lower frequency
cut-off is imposed by seismic noise which sharply reduces sensitivity at
low frequencies,
while the upper cut-off corresponds roughly to the frequency at which the
sensitivity degrades to the level found at the low frequency cut-off.
(A dedicated search for bursts with frequency content above 2000\,Hz is
presented in a companion paper~\cite{S5burstAllSkyH}.)

\subsection{Event-by-event DQFs and vetoes}

After gravitational-wave triggers have been identified by an analysis
pipeline, they are checked against additional DQFs and ``veto''
conditions to see if they occurred within a time interval which should
be excluded from the search.  The DQFs applied at this stage consist
of many short intervals which would have fragmented the data set if
applied in the initial data selection stage.
Event-by-event veto conditions are based on a
statistical correlation between the rate of transients in
the GW channel and noise transients, 
or ``glitches'', in environmental and  interferometric auxiliary channels. 
The performance of vetoes (as well as DQFs) are evaluated by the extent to which they remove the GW channel transients
of each interferometer, as identified by the KleineWelle (KW)~\cite{blackburn2005} algorithm. KW looks for
excess signal energy by decomposing a timeseries into the Haar wavelet domain. 
For each transient, KW calculates a significance defined as the negative of the 
natural logarithm of the probability, in Gaussian noise, of observing an event as energetic or more than the one in consideration.
The veto conditions, like the DQFs, were completely defined before
unshifted data was analyzed to identify gravitational-wave event
candidates.
A detailed description of the implementation of the vetoes is given in 
Appendix B.

\subsection{Background estimation}

In order to estimate the false trigger rate from detector noise fluctuations
and artifacts, data from the various detectors
are artificially shifted in time so as to remove any coincident signals.
These time-shifts have strides much longer than the intersite time-of-flight
for a true gravitational-wave signal and thus are unlikely to preserve any
reconstructable astrophysical signal when analyzed.
We refer to these as time-shifted data.
Both unshifted and time-shifted data are analyzed by identical
procedures, yielding the candidate sample and the estimated background
of the search, respectively.
In order to avoid any biases, no unshifted data are used
in the tuning of the methods.
Instead, combined with simulations (see below), background data
are used as the test set over which all analysis cuts are defined
{\bf{prior}} to examining the unshifted data-set.
In this way, our analyses are ``blind''.

\subsection{Hardware signal injections}

During the S5 run, simulated
GW signals were occasionally injected into the data by applying
an actuation to the mirrors at the ends of the interferometer
arms.  The waveforms and times of the injections were cataloged for
later study.
These were analyzed as an end-to-end validation of the
interferometer readout, calibration, and detection algorithms.

\subsection{Simulations}

In addition to analyzing the recorded data
stream in its original form, many simulated
signals are injected
in software---by adding the signal to the digital data stream---in
order to to simulate the passage of
gravitational-wave bursts through the network of detectors.
The same simulated signals are analyzed by all three analysis pipelines.
This provides a means for establishing the sensitivity of the
search by measuring the probability of detection as a function
of the signal morphology and strength.
These will also be referred to as efficiency curves.

\section{Search Algorithms}
\label{sec:algorithms}

Unmodeled GW bursts can be distinguished from instrumental noise if they 
show consistency in time, frequency, shape, and amplitude among the   
LIGO detectors. The time constraints, for example, follow from the maximum possible 
propagation delay between the Hanford and Livingston sites which is  10\,ms. 

This S5 analysis employs three algorithms to search for GW bursts: BlockNormal \cite{finn04},
QPipeline~\cite{chatterjiThesis,multiresolution}, and coherent WaveBurst~\cite{Klimenko:2008fu}. A detailed description of each algorithm can be found in the
appendices. Here we limit ourselves to a brief summary of the three techniques. All three
algorithms essentially look for
excess power~\cite{anderson01} in a time-frequency decomposition
of the data stream. Events are ranked and checked for temporal coincidence and coherence
(defined differently for the different algorithms) across the network of detectors.
The three techniques differ in the details of how the time-frequency decompositions are performed,
how the excess power is computed, and how coherence is assessed. 
Each analysis pipeline was independently developed, coded and tuned.
Because the three pipelines have
different sensitivities to different types of GW signals and instrumental artifacts, the
results of the three searches can be combined to produce stronger statements about event candidates and
upper limits. 

BlockNormal (BN) performs a time-frequency decomposition by taking short segments of data and applying a heterodyne
basebanding procedure to divide each segment into frequency bands. A change-point analysis is used to identify events
with excess power in each frequency band for each detector, and events are clustered to form single-interferometer
triggers. Triggers from the various interferometers that fall within a certain coincidence window are then combined
to compute the ``combined power'', $P_{\text{C}}$, across the network. These coincident triggers are then checked for coherence
using CorrPower, which calculates a cross-correlation statistic $\Gamma$ that
was also used in the S4 search~\cite{S4burstAllSky}. A detailed description
of the BN algorithm can be found in Appendix~\ref{sec:blocknormal}.

QPipeline (QP) performs a time-frequency decomposition by filtering
the data against bisquare-enveloped sine waves, in what
amounts to an over-sampled wavelet transform. The filtering procedure yields a standard matched filter signal to noise ratio (SNR), $\rho$,
which is used to identify excess power events in each interferometer (quoted in terms of the quantity $Z=\rho^2/2$).
Triggers from the various interferometers are combined to give candidate events if they have consistent central times
and frequencies. QPipeline also looks for coherence in the response of the H1 and H2 interferometers by comparing
the excess power of sums (the coherent combination $\text{H}+$) and differences (the null combination $\text{H}-$) of the data. Rather than using the single-interferometer
H1, H2, L1, signal to noise ratios, the QPipeline analysis uses the SNRs in the transformed
channels $\text{H}+$, $\text{H}-$, and  L1.
A detailed description of the QPipeline algorithm can be found in Appendix~\ref{sec:qpipeline}.

Coherent WaveBurst (cWB) performs a time-frequency decomposition using critically sampled Meyer wavelets. 
The cWB version used in S5 replaces the separate coincidence and correlation test (CorrPower) used in the S4 analysis~\cite{S4burstAllSky} by a single
coherent search statistic based on a Gaussian likelihood function. Constrained waveform reconstruction is used to compute the {\em network} likelihood and
a coherent network amplitude. This coherent analysis has
the advantage that it is not limited by the performance of the least sensitive detector in the network.
In the cWB analysis, various signal combinations are used to measure the signal consistency among different sites: a network correlation
statistic $cc$
, network energy disbalance $\Lambda_{\text{NET}}$, H1-H2 disbalance $\Lambda_{\text{HH}}$
and a penalty factor $P_f$.
These quantities are used in concert with the coherent network amplitude $\eta$
to develop efficient selection cuts that can eliminate spurious events with a very limited impact on the sensitivity.
It is worth noting that the version of cWB used in the S5 search is 
more advanced than the one used on LIGO and GEO data in S4~\cite{S4LIGOGEO}. 
A detailed description of the cWB algorithm can be found in Appendix~\ref{sec:cWB}.

Both QPipeline and coherent WaveBurst use
the freedom to form linear combinations of the data to construct ``null streams'' that are insensitive to GWs. These null streams provide a powerful tool for distinguishing between genuine GW signals
and instrument artifacts~\cite{shourov2006}.

\section{Background and tuning}\label{sec:tuning}

As mentioned in Sec.~IV,
the statistical properties of the noise triggers (background) 
are studied for all network combinations
by analyzing time-shifted data,
while the detection capabilities of the search pipelines for various
types of GW signals are studied by analyzing simulated signals
(described in the following section) injected into actual detector noise.
Plots of the parameters for noise triggers and signal injections
are then examined to tune the searches.
Thresholds on the parameters are chosen to maximize the efficiency in detecting GWs
for a predetermined, conservative false alarm rate of roughly 5 events for every
100 time shifts of the full data set, {\it i.e.}\ $\sim$$0.05$ events
expected for the duration of the data set.

For a given energy threshold, all three pipelines observed a much
larger rate of triggers with frequencies below 200\,Hz than at higher
frequencies.  Therefore, each pipeline set separate thresholds for
triggers above and below 200\,Hz, maintaining good sensitivity for
higher-frequency signals at the expense of some sensitivity for
low-frequency signals.  The thresholds were tuned separately for each
detector network, and the cWB pipeline also distinguished among a few
distinct epochs with different noise properties during the run.
A more detailed description of the tuning process can be found in Appendices C, D, and E.

\section{Simulated signals and efficiency curves}\label{sec:simulations}
In this section we present the efficiencies of the different algorithms 
in detecting simulated GWs.
As in previous science runs, we do not attempt to survey the complete spectrum 
of astrophysically motivated signals.  Instead, we use a limited number 
of ad hoc waveforms that probe the range of frequencies of interest, 
different signal durations, and different GW polarizations.

We choose three families of waveforms: sine-Gaussians, Gaussians, and 
``white noise bursts''.
An isotropic sky distribution was generated in all cases.
The Gaussian and sine-Gaussian signals
have a uniformly distributed random linear
polarization, while the white noise bursts contain approximately
equal power in both polarizations.
We define the amplitude of an injection in terms of the total signal energy at
the Earth observable by an ideal optimally oriented detector able
to independently measure both signal polarizations:
\begin{eqnarray}
h_{\text{rss}}^2 & = & {\int_{-\infty}^{+\infty} ( |h_+(t)|^2 + |h_{\times}(t)|^2 ) \, dt}\\ \nonumber
  &  =  &{\int_{-\infty}^{+\infty} ( |\tilde{h}_+(f)|^2 + |\tilde{h}_{\times}(f)|^2 ) \, df.}
\end{eqnarray}

In reality, the signal observed at an individual detector depends on the
direction $\hat{\Omega}$ to the source and the polarization angle $\Psi$ 
through ``antenna factors'' $F_+$ and $F_\times$:
\begin{equation}
\label{eq:hdet}
h_{\text{det}} = F_+(\hat{\Omega}, \Psi) h_+ + F_{\times}(\hat{\Omega}, \Psi) h_{\times} \, .
\end{equation}

In order to estimate the detection efficiency as a function of signal strength,
the simulated signals were injected at 22 logarithmically spaced
values of $h_{\text{rss}}$ ranging from
$1.3 \times 10^{-22} \, \mathrm{Hz}^{-1/2} $ to
$1.8 \times 10^{-19} \, \mathrm{Hz}^{-1/2} $\,, stepping
by 
factors of $\sim$$\surd{2}$. Injections were performed at quasi-random times regardless of data quality or
detector state, with an average rate of one injection every 100 seconds.
The efficiency of a method is then defined as the fraction 
of waveforms that are detected out of all that were injected into the
data analyzed by the method.

\subsection{Simulated signals}

The first family of injected signals are sine-Gaussians. These are sinusoids with a central frequency $f_0$, dimensionless width $Q$ and arrival time $t_0$,
defined by:
\begin{equation} 
h_+(t_0+t)=h_0 \sin(2\pi f_0 t)\exp[-(2\pi f_0 t)^2/2Q^2].
\end{equation}
More specifically $f_0$ was chosen to be one of
(70, 100, 153, 235, 361, 554, 849, 945, 1053, 1172, 1304, 1451, 1615, 1797, 2000) \,Hz; and
 $Q$ to be one of 3, 9, or 100.

The second family consists of Gaussian pulses described by 
the following expression:
\begin{equation}
h_+(t_0 + t ) = h_0 \exp(-t^2 /\tau^2 )
\end{equation}
where $\tau$ is chosen to be one of (0.05, 0.1, 0.25, 0.5, 1.0, 2.5, 4.0, 6.0, 8.0) ms.

The third family are the ``white noise bursts'' (WNBs).
These were generated by bandpassing white noise in frequency bands
starting at 100\,Hz, 250\,Hz, or  1000\,Hz, 
with bandwidth  10\,Hz,  100\,Hz, or 1000\,Hz, and by time windowing with Gaussian 
profiles of duration (half of the interval between the inflection points) equal to 100\,ms, 10\,ms, or 1\,ms.
For each waveform type (a choice of central frequency, bandwidth, and duration), 30 waveform files with random data
content were created. 
The injections for each waveform type use random pairs selected 
from the 30 created waveforms for the $h_+$ and $h_\times$ polarizations
(the selection avoids pairs with identical waveforms). 
This results in unpolarized injections with equal amounts of power on average in
each polarization state.

\begin{figure}
\raisebox{1.8in}{(a)}\includegraphics[width=3in]{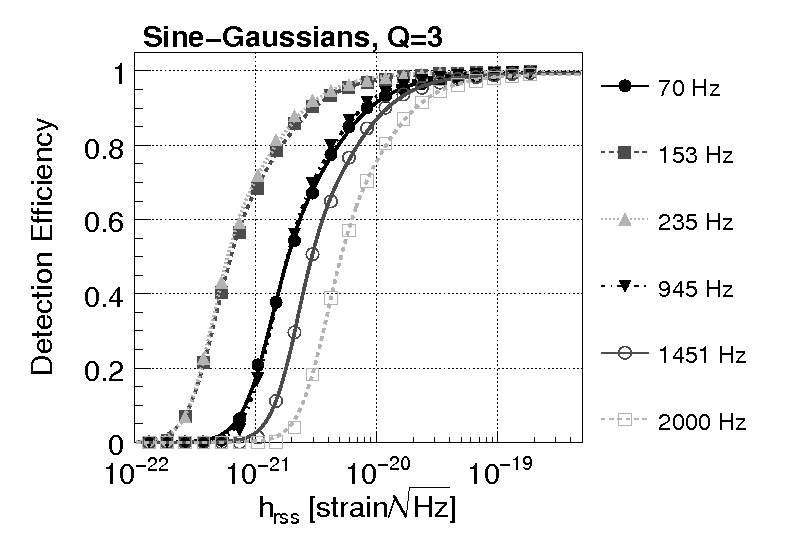}\\
\raisebox{1.8in}{(b)}\includegraphics[width=3in]{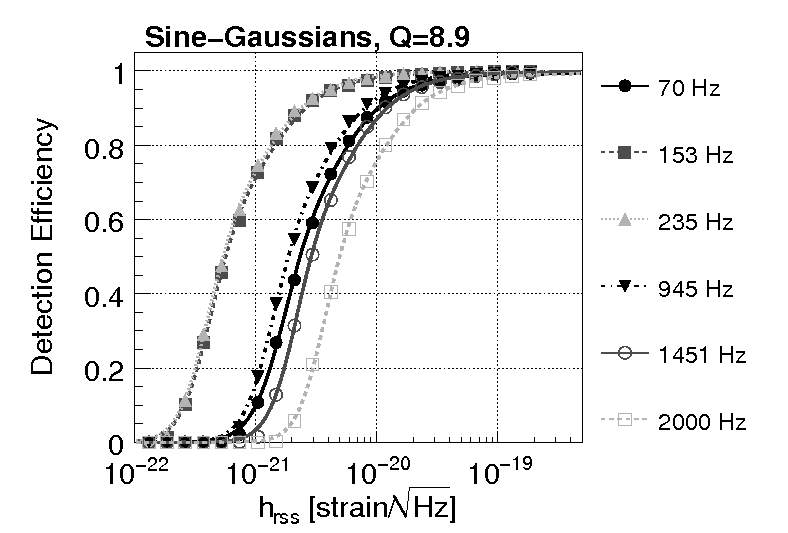}\\
\raisebox{1.8in}{(c)}\includegraphics[width=3in]{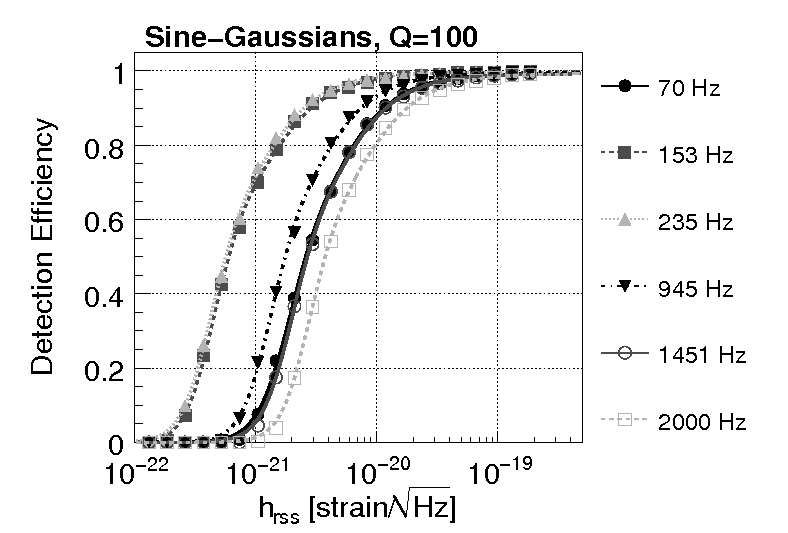}
\caption{\label{figure:efficiencies_combined_1}
Combined efficiencies of the three pipelines and two networks (H1H2L1
and H1H2) used in the upper limit analysis for selected sine-Gaussian waveforms with (a) $Q=3$, (b) $Q=9$, (c) $Q=100$.
 These efficiencies have been calculated using the logical OR of the
pipelines and networks for the subset of simulated signals that
were injected in time intervals that were actually analyzed, and thus
approach unity for large amplitudes.
}
\end{figure}

\begin{figure}
\raisebox{1.8in}{(a)}\includegraphics[width=3in]{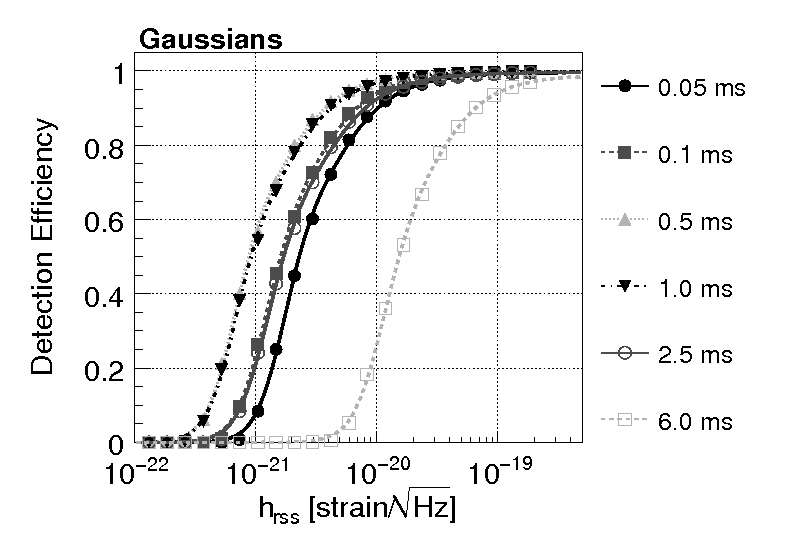}\\
\raisebox{1.8in}{(b)}\includegraphics[width=3in]{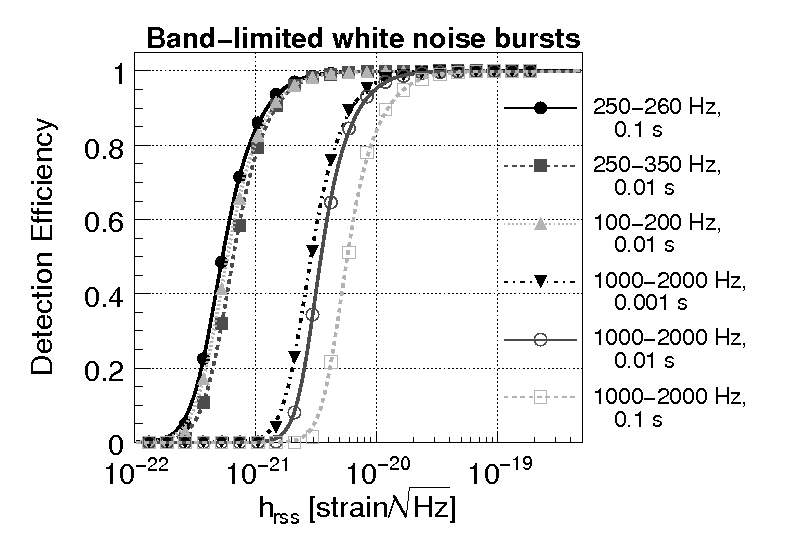}
\caption{\label{figure:efficiencies_combined_2}
Combined efficiency of the three pipelines and two networks (H1H2L1
and H1H2) used in the upper limit analysis for (a) selected linearly-polarized
Gaussian waveforms; (b) selected band-limited white-noise bursts with two independent
polarization components.
 These efficiencies have been calculated using the logical OR of the
pipelines and networks for the subset of simulated signals that
were injected in time intervals that were actually analyzed, and thus
approach unity for large amplitudes.
}
\end{figure}

Each efficiency curve, consisting of the efficiencies determined for
a given signal morphology at each of the 22 $h_{\text{rss}}$ values,
was fitted with an empirical four-parameter function.
The efficiency curves for the logical OR combination of the three
pipelines and for the combined H1H2 and H1H2L1 networks
are shown for selected waveforms in Figs.~\ref{figure:efficiencies_combined_1} and~\ref{figure:efficiencies_combined_2}.
The $h_{\text{rss}}$ values yielding 50\% detection efficiency,
 $h_{\mathrm{rss}}^{50\%}$,
are shown in Tables I and II for sine-Gaussians with $Q=9$
and for white noise bursts injected and analyzed in H1H2L1 data.
The study of the efficiency for all the waveforms shows that the combination of the methods is slightly 
more sensitive than the best performing one, which is QPipeline for some of the
sine-Gaussians, and cWB for all other waveforms considered.

\begin{table}
  \centering
  \begin{tabular}{rrrrrr}\hline\hline
$f$ (Hz) & $Q$ &  Combined  &  cWB   &  BN    & QP  \\ \hline
  70   &  9 &  25.8 &  25.9 & 227.4  & 33.1 \\
 100   &  9 &  10.3 &  10.5 &  13.6  & 14.0 \\
 153   &  9 &   6.3 &   6.5 &   7.8  &  8.8 \\
 235   &  9 &   6.0 &   6.3 &   7.7  &  6.8 \\
 361   &  9 &  10.9 &  11.2 &  16.3  & 12.0 \\
 554   &  9 &  12.0 &  12.6 &  15.5  & 12.9 \\
 849   &  9 &  18.1 &  19.0 &  23.7  & 19.2 \\
   945   &  9 &  20.6 &  21.6 &  27.8  & 22.2 \\
  1053   &  9 &  23.3 &  24.8 &  33.4  & 24.1 \\
  1172   &  9 &  25.2 &  26.8 &  36.5  & 26.3 \\
  1304   &  9 &  28.7 &  30.9 &  40.8  & 29.5 \\
  1451   &  9 &  32.0 &  35.0 &  48.1  & 32.9 \\
  1615   &  9 &  35.2 &  38.2 &  51.5  & 36.3 \\
  1797   &  9 &  42.0 &  44.2 &  62.2  & 45.4 \\
  2000   &  9 &  54.5 &  55.9 &  77.6  & 68.8 \\
                \hline\hline
                \end{tabular}
        \caption{$h_{\text{rss}}$ values yielding 50\% detection efficiency,
     in units of $10^{-22} \, {\rm Hz}^{-1/2}$,
     for different sine-Gaussian waveforms and pipelines
in the H1H2L1 network.  The first column is the central frequency, the second the quality factor, the third 
the $h_{\mathrm{rss}}^{50\%}$ of the logical OR of the pipelines, and the remaining three columns the $h_{\mathrm{rss}}^{50\%}$ of the individual pipelines.
These $h_{rss}^{50\%}$ values include an adjustment of 11.1\% to take
into account calibration and statistical uncertainties as explained in
Sec.~\ref{sec:systematics}.}
        \label{table:jjj}
\end{table}

\begin{table}
  \centering
  \begin{tabular}{rrrrrrr}\hline\hline
$f$ (Hz)& BW (Hz) & $d$ (ms) & Combined &   cWB &   BN   & QP  \\ \hline
1000   & 1000   & 0.001   & 32.0   & 34.4   & 51.8   & 33.2 \\
1000   & 1000   & 0.01    & 38.6   & 39.1   & 47.1   & 51.9 \\
1000   & 1000   & 0.1     & 63.4   & 65.8   & 73.0   & 113.6 \\
1000   & 100    & 0.01    & 22.2   & 22.6   & 30.9   & 25.9 \\
1000   & 100    & 0.1     & 28.5   & 28.5   & 44.6   & 44.6 \\
1000   & 10     & 0.1     & 21.5   & 21.4   & 30.8   & 44.8 \\
100    & 100    & 0.01    & 6.5    & 6.7    & 7.5    & 9.2 \\
100    & 100    & 0.1     & 7.9    & 7.9    & 9.9    & 14.1 \\
100    & 10     & 0.1     & 9.1    & 9.1    & 13.7   & 12.7 \\
250    & 100    & 0.01    & 7.3    & 7.6    & 18.6   & 8.5 \\
250    & 100    & 0.1     & 8.8    & 8.9    & 11.6   & 13.4 \\
250    & 10     & 0.1     & 5.9    & 5.9    & 9.0    & 17.6 \\
                \hline\hline
                \end{tabular}
        \caption{$h_{\text{rss}}$ values yielding 50\% detection efficiency,
             in units of $10^{-22} \, {\rm Hz}^{-1/2}$,
             for different white noise burst  waveforms and pipelines
in the H1H2L1 network.  The first column is the central frequency, the second the bandwidth, the third the duration of the 
gaussian window, the fourth 
the $h_{\mathrm{rss}}^{50\%}$ of the logical OR of the pipelines, and the remaining three columns the $h_{\mathrm{rss}}^{50\%}$ of the individual pipelines.
These $h_{rss}^{50\%}$ values include an adjustment of 11.1\% to take
into account calibration and statistical uncertainties as explained in
Sec.~\ref{sec:systematics}.}

        \label{table:jj}
\end{table}

\section{Statistical and calibration errors}\label{sec:systematics}

The $h_{\text{rss}}^{50\%}$ values presented in this paper have been
adjusted to conservatively reflect
systematic and statistical uncertainties.
The dominant source of systematic uncertainty
is from the amplitude measurements in the frequency domain calibration. 
The individual amplitude uncertainties from each interferometer can be 
combined into a single uncertainty by calculating a combined root-sum-square
amplitude SNR and propagating the individual uncertainties
assuming each error is independent. 
In addition, there is a small uncertainty (about 1\%) 
introduced by converting from the frequency domain to the
time domain strain series on which the analysis was actually run.  
There is also phase uncertainty on the order of a few degrees in each
interferometer, arising both from the initial
frequency domain calibration and the conversion to the time domain.
However, this is not a significant concern since the phase
uncertainties at all frequencies correspond to phase shifts on the order
of less than half a sample duration.  We therefore do not make any adjustment
to the overall systematic uncertainties due to phase error.
Finally, statistical uncertainties on the fit parameters (arising
from the binomial errors on the efficiency measurements)
affect $h_{\text{rss}}^{50\%}$ by approximately 1.4\% 
on average and are not much different for any particular waveform.  

The frequency-domain amplitude uncertainties are added in quadrature 
with the other smaller uncertainties to obtain a total 1-sigma relative 
error for the SNR.  The relative error in the 
$h_{\text{rss}}$ is then the same as the relative error in the SNR.
Thus, we adjust our sensitivity estimates by increasing 
the $h_{\text{rss}}^{50\%}$ values by the reported percent uncertainties multiplied by 1.28 
(to rescale from a 1-sigma fluctuation to a 90\% confidence level 
upper limit, assuming Gaussian behavior), which amounts to 11.1\% in the
frequency band explored in this paper.

\section{Search Results}\label{sec:results}

\begin{figure}[htbp]
\begin{center}
 \includegraphics[height=5cm]{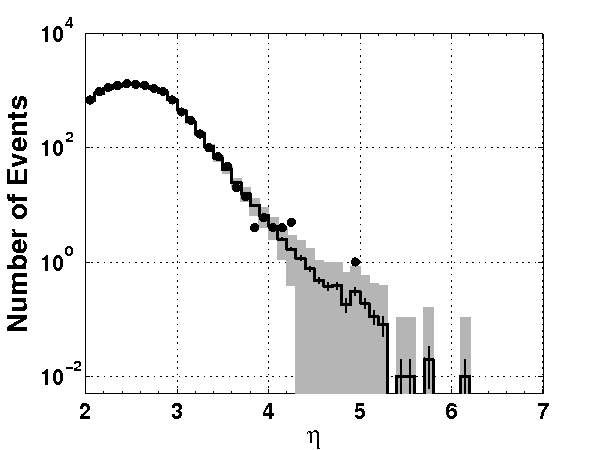}\\
 \includegraphics[height=5cm]{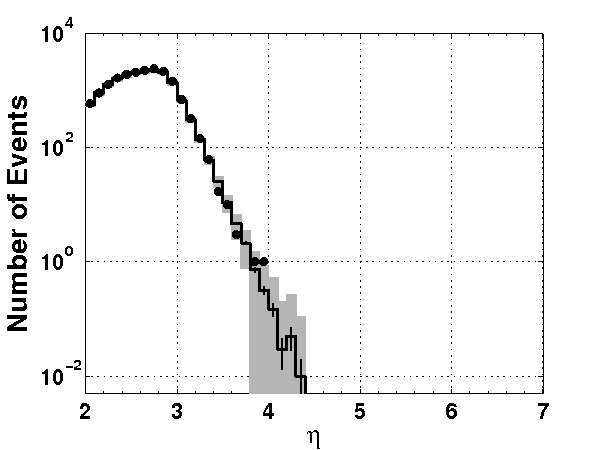}\\
\caption{\small Distributions of cWB H1H2L1 triggers after category 2
DQFs were applied. Overlaid histograms for $\eta$ for unshifted
triggers (dots) and mean background estimated from time-shifted
triggers (stair-step curve).
The narrow error bars indicate the statistical uncertainty of the
background estimate, while the shaded band indicates the expected
root-mean-square statistical fluctuations on the number of background
triggers in each bin.  The top panel represents the triggers with
central frequency below 200\,Hz while the bottom panel represents the
triggers with central frequency above 200\,Hz.}
\label{fg:histCWB}
\end{center}
\end{figure}

\begin{figure}[htbp]
\begin{center}
 \includegraphics[height=5.2cm]{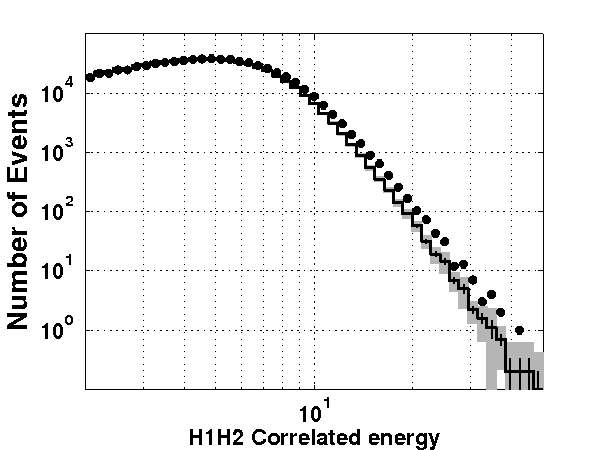}\\
 \includegraphics[height=5.2cm]{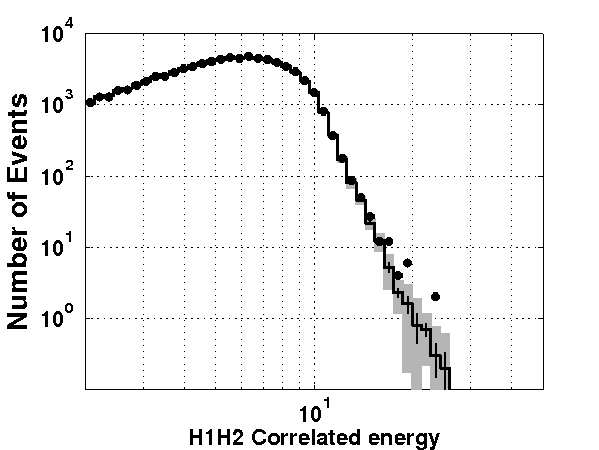}
\caption{\small QPipeline triggers after category 2 DQFs were
applied. Overlaid histograms for H1H2 correlated energy for unshifted
H1H2 triggers (dots) and mean background estimated from time-shifted
triggers (stair-step curve).
The narrow error bars indicate the statistical uncertainty of the
background estimate, while the shaded band indicates the expected
root-mean-square statistical fluctuations on the number of background
triggers in each bin.  The top panel represents the triggers with
central frequency below 200\,Hz while the bottom panel represents the
triggers with central frequency above 200\,Hz.}
\label{fg:histQ}
\end{center}
\end{figure}

\begin{figure}[htbp]
\begin{center}
 \includegraphics[width=2.75in]{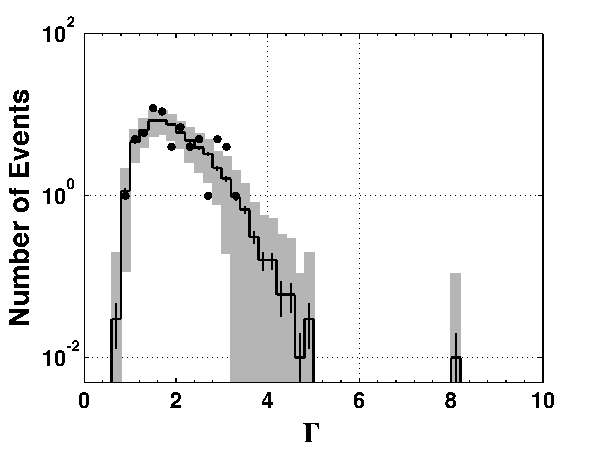}\\
 \includegraphics[width=2.75in]{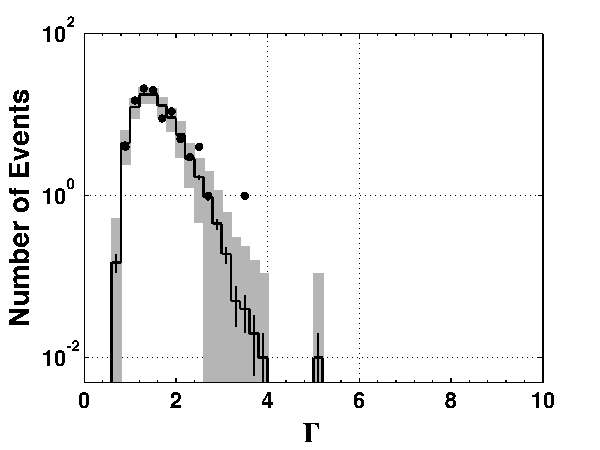}
\caption{\small BlockNormal triggers after category 2 DQFs were
applied. Overlaid histograms for $\Gamma$ for unshifted H1H2L1
triggers (dots) and mean background estimated from time-shifted
triggers (stair-step curve).
The narrow error bars indicate the statistical uncertainty of the
background estimate, while the shaded band indicates the expected
root-mean-square statistical fluctuations on the number of background
triggers in each bin.  The top panel represents the triggers with
central frequency below 200\,Hz while the bottom panel represents the
triggers with central frequency above 200\,Hz.}
\label{fg:histBN}
\end{center}
\end{figure}

Once category 2 DQFs have been applied on the triggers produced from the
unshifted ({\it i.e.}\ candidate sample) and time-shifted (background) data, 
histograms of the two populations are generated 
for each pipeline, interferometer network and frequency band.
See for example trigger distributions for the H1H2L1 network
in Figs.~\ref{fg:histCWB}, \ref{fg:histQ}, and \ref{fg:histBN}.
No unshifted triggers are found above threshold in the final sample 
for any of the three pipelines and four network configurations.
We therefore have no candidate GW signals, and no follow up for 
possible detections is performed.  We proceed to set upper limits 
on the rate of specific classes of GWs.


\subsection{Upper limits}\label{sec:upper_limits}

Our measurements consist of the list of triggers detected by 
each analysis pipeline (BN, QP, cWB) in each network data set 
(H1H2L1, H1H2, H1L1, H2L1).  BN analyzed the H1H2L1 data, 
QP analyzed H1H2L1 and H1H2, and cWB analyzed all four data 
sets.  In general, the contribution to the upper limit due 
to a given pipeline and data set increases with both the 
detection efficiency of the pipeline and the livetime of 
the data set.  Since the duty cycle of the H1L1 and H2L1 data 
sets is small (2.4\% and 4.5\% after category 3 DQFs and
 category 3 vetoes, 
vs.~37.2\% and 22.5\% in H1H2L1 and H1H2), and the data 
quality not as good, we decided {\it a priori} to not 
include these data sets in the upper limit calculation.
We are therefore left with five analysis pipeline results: 
BN-H1H2L1, QP-H1H2L1, QP-H1H2, cWB-H1H2L1, and cWB-H1H2. 
We wish to combine these 5 results to produce a single 
upper limit on the rate of GW bursts of each of the 
morphologies tested.

We use the approach described in \cite{sutton2009} 
to combine the results of the different search detection 
algorithms and networks. Here we give only a brief summary 
of the technique.

The procedure given in~\cite{sutton2009} is to combine the 
sets of triggers according to which pipeline(s) and/or network
detected any given trigger.  For example, in the case 
of two pipelines ``A'' and ``B'', the outcome of the counting 
experiment is the set of three numbers $\vec{n}=(n_{\text{A}}, n_{\text{B}}, n_{\text{AB}})$, 
where $n_{\text{A}}$ is the number of events detected by pipeline A but not 
by B, $n_{\text{B}}$ is the number detected by B but not by A, and $n_{\text{AB}}$ 
is the number detected by both.  (The extension to an 
arbitrary number of pipelines and data sets is straightforward.)
Similarly, one 
characterizes the sensitivity of the experiment by the probability 
that any given GW burst will be detected by a given combination of 
pipelines.  We therefore compute the efficiencies 
$\vec{\epsilon}=(\epsilon_{\text{A}},\epsilon_{\text{B}},\epsilon_{\text{AB}})$, where 
$\epsilon_{\text{A}}$ is the fraction of GW injections that are detected 
by pipeline A but not by B, etc.

To set an upper limit, one must decide {\it a priori} how to 
rank all possible observations, so as to determine whether a given
observation $\vec{n}$ contains ``more'' or ``fewer'' events than 
some other observation $\vec{n}'$.  Denote the ranking function 
by $\zeta(\vec{n})$.  Once this choice is made, the actual 
set of unshifted events is observed, giving $\vec{n}$, and the 
rate upper limit $R_{\alpha}$ at confidence level $\alpha$ is 
given by  
\begin{equation}\label{eqn:ulk}
  1 - \alpha = \sum_{\vec{N}|\zeta(\vec{N}) \le \zeta(\vec{n})} P(\vec{N}|\vec{\epsilon},R_{\alpha} \vec{T}) \, . 
\end{equation}
Here $P(\vec{N}|\vec{\epsilon},R_\alpha \vec{T})$ is the prior 
probability of observing $\vec{N}$ given the true GW rate 
$R_\alpha$, the vector containing the livetimes of different data sets $\vec{T}$
(this is a scalar if we are combining results of methods analyzing the 
same livetime), and the detection efficiencies 
$\vec{\epsilon}$.  The sum is taken over all $\vec{N}$ for which 
$\zeta(\vec{N}) \le \zeta(\vec{n})$; i.e., over all 
possible outcomes $\vec{N}$ that result in ``as few or fewer'' 
events than were actually observed.

As shown in~\cite{sutton2009}, a convenient choice for 
the rank ordering is 
\begin{equation}\label{eqn:rank}
  \zeta(\vec{n}) = \vec{\epsilon}\cdot\vec{n}  \, .
\end{equation}
That is, we weight the individual measurements $(n_{\text{A}},n_{\text{B}},n_{\text{AB}},\ldots)$ 
proportionally to the corresponding efficiency 
$(\epsilon_{\text{A}},\epsilon_{\text{B}},\epsilon_{\text{AB}},\ldots)$. 
This simple procedure yields a single upper limit from the 
multiple measurements.  From the practical point of view, 
it has the useful properties that the pipelines need not be 
independent, and that combinations of pipelines and data sets in 
which it is less likely for a signal to appear (relatively 
low $\epsilon_i$) are naturally given less weight.

Note that for the purpose of computing the upper limit on 
the GW, we are ignoring any background.  This leads to our 
limits being somewhat conservative, since a non-zero background 
contribution to $\vec{n}$ will tend to increase the estimated limit.

\begin{figure}
\includegraphics[width=3.0in, viewport= 50 0 790 450, clip=true]{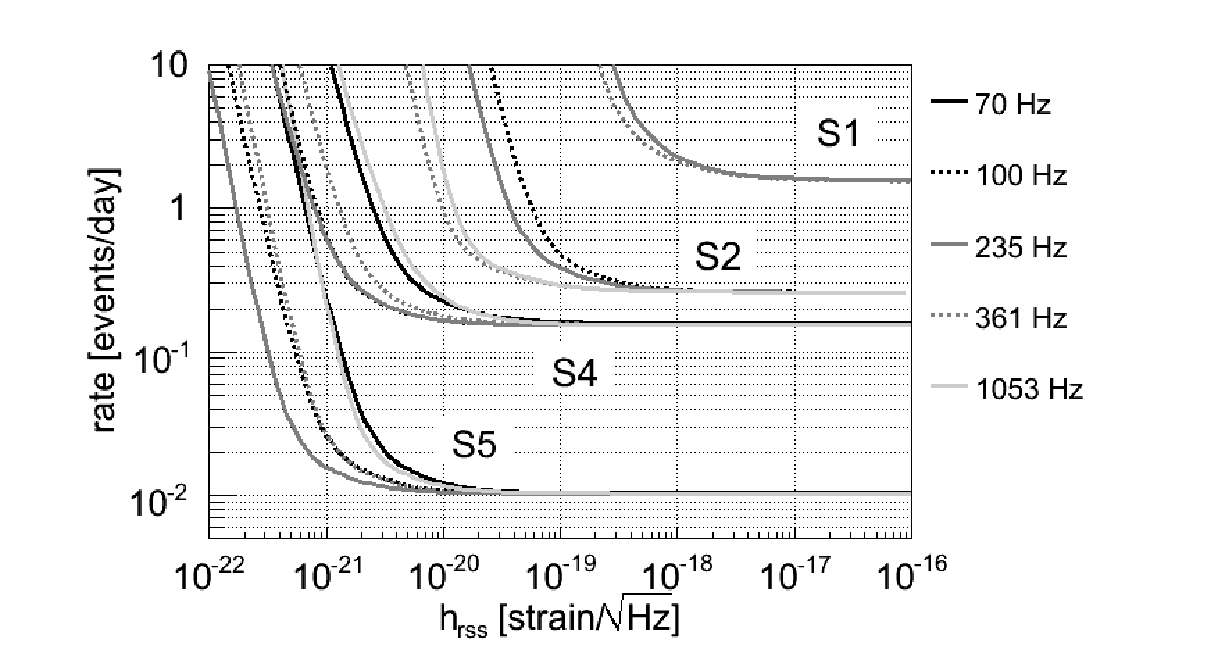}\\
\includegraphics[width=3.0in, viewport= 50 0 790 450, clip=true]{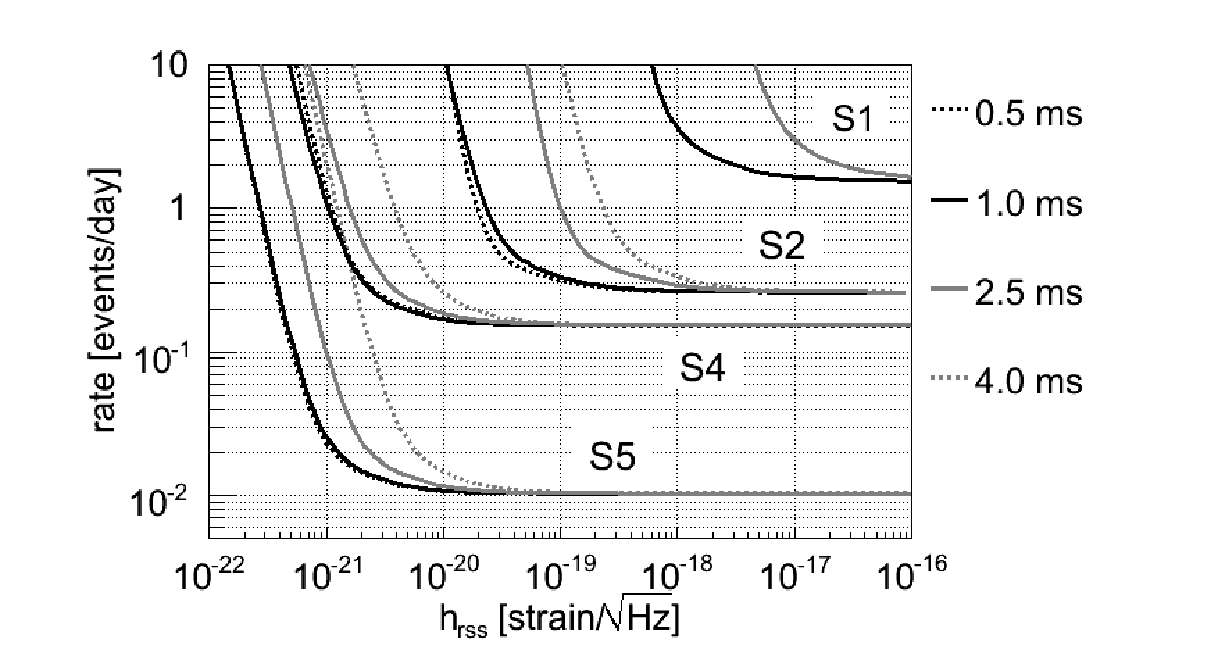}
\caption{Selected exclusion diagrams showing the 90\% confidence rate limit as a function
of signal amplitude for Q=9 sine-Gaussian (top) and Gaussian (bottom) waveforms
for the results in this paper (S5) compared to the results reported previously
(S1, S2, and S4).}
\label{figure:upperlimits_compare_ga2}
\end{figure}

In the present search, no events were detected by any analysis 
pipeline, so $\vec{n}=\vec{0}$.  As shown in~\cite{sutton2009}, 
in this case the efficiency weighted upper limit procedure given by
Eqs.~(\ref{eqn:ulk}) and~(\ref{eqn:rank}) gives a particularly simple 
result: the procedure is equivalent to taking the logical OR of all
five pipeline/network samples.
The $\alpha=90\%$ confidence level upper limit 
for zero observed events, $R_{90\%}$, is given by
\begin{eqnarray}\label{eqn:spul0}
0.1 & = & \exp(-\epsilon_{\text{tot}}R_{90\%}T) \\
\Rightarrow R_{90\%} & = & \frac{2.30}{\epsilon_{\text{tot}}T} \, ,
\end{eqnarray}
where $\epsilon_{\text{tot}}$ is the 
weighted average of all the efficiencies (the weight is the 
relative livetime) and $T$ is the total observation time.
Fig.~\ref{figure:upperlimits_compare_ga2} shows the combined rate 
upper limits as a function of amplitude for selected sine-Gaussian and 
Gaussian GW bursts.  
 In the limit of strong signals, 
$\epsilon_{\text{tot}}T$ goes to 224.0 days which is the union of all time analyzed 
for the H1H2L1 and H1H2 networks after category 3 DQFs.
The rate limit thus becomes 
$0.0103\,\textrm{day}^{-1}=3.75\,\textrm{yr}^{-1}$.

\section{Summary and Discussion}\label{sec:summaries}

The search for unmodeled GW bursts reported in this paper is currently the most sensitive
ever performed.
The quality of the data and the sensitivity of the data analysis algorithms have improved since
the S4 run, and the quantity of data available for analysis has increased by more than an order of magnitude. These improvements
are reflected in the greater strain sensitivity (with $h_{\text{rss}^{50\%}}$
values as low as $\sim 6\times 10^{-22}\,\text{Hz}^{-1/2}$) and the
tighter limit on the rate of bursts
(less than 3.75 events per year at 90\% confidence level)
with large enough amplitudes to be detected reliably.
The most sensitive previous search, using LIGO S4 data, achieved
$h_{\text{rss}^{50\%}}$ sensitivites as low as a few times
$10^{-21} \,\text{Hz}^{-1/2}$ and a rate limit of 55 events per year.
We note that the IGEC network of resonant bar detectors has set a more
stringent rate limit, $1.5$ events per year at 95\% confidence
level~\cite{astone2003}, for GW bursts near the resonant frequencies
of the bars with $h_{\text{rss}} > \sim 8 \times 10^{-19}
\,\text{Hz}^{-1/2}$ (see Sec.\ X of~\cite{S2burstAllSky} for the
details of this comparison).  A later joint observation run,
IGEC-2, was a factor of $\sim$3 more sensitive but had shorter
observation time~\cite{astone2007}.

In order to set an astrophysical scale to the sensitivity achieved by this search,
we now repeat the analysis and the examples presented
for S4.
Specifically, we can estimate what amount of mass converted into GW burst 
energy at a given distance would be strong enough to be detected by the search
with 50\% efficiency. Following the same steps as in~\cite{S4burstAllSky}, 
assuming isotropic emission and a distance of 10\,kpc we  
find that a 153\,Hz sine-Gaussian with $Q=9$ would need $1.9 \times 10^{-8}$ solar masses, 
while for S4 the figure was $10^{-7}\,M_\odot$.
For a source in the Virgo galaxy cluster, approximately 16\,Mpc away, the 
same $h_{\text{rss}}$ would be produced by an energy emission of roughly 
$0.05\,M_\odot c^2$, while for S4 it was $0.25\,M_\odot c^2$.

We can also update our estimates for the detectability of two classes 
of astrophysical sources: core collapse supernovae and binary 
black-hole mergers.
We consider first the core collapse supernova simulations by Ott.\ et al.~\cite{ott06}.
In this paper gravitational waveforms were computed for three progenitor 
models: s11WW, m15b6 and s25WW.
From S4 to S5 the astrophysical reach 
for the s11WW and m15b6 models
improved from approximately 0.2 to 0.6 kpc while for s25WW it improved from 8 to 24 kpc.
Second, we consider the binary black hole merger  
calculated by the Goddard 
numerical relativity group~\cite{baker2006}.
A binary system of two 10-solar-mass black holes (total $20\,M_\odot$) would 
be detectable with 50\% efficiency at a distance of roughly 4\,Mpc
compared to 1.4\,Mpc in S4, 
while a system with total mass $100\,M_\odot$ 
would be detectable out to $\sim$180\,Mpc, compared to $\sim$60\,Mpc in S4.
In each case the astrophysical reach has improved by approximately a
factor of 3 from S4 to S5.

At present, the analysis of the second year of S5 is well underway, including a joint analysis of
data from Virgo's VSR1 run which overlaps with the final 4.5 months of S5. Along with the potential for
better sky coverage, position reconstruction and glitch rejection, the joint analysis brings with it new
challenges and opportunities.
Looking further ahead, the sixth LIGO science run and second Virgo science
run are scheduled to start in mid 2009, with the two LIGO 4\,km
interferometers operating in an ``enhanced'' configuration that is
aimed at delivering approximately a factor of two improvement in
sensitivity, and comparable improvements for Virgo.
Thus we will soon be able to search for GW bursts farther out into the
universe.

\begin{acknowledgments}\label{sec:acknowledgements}

The authors gratefully acknowledge the support of the United States National
Science Foundation for the construction and operation of the LIGO Laboratory and
the Science and Technology Facilities Council of the United Kingdom, the
Max-Planck-Society, and the State of Niedersachsen in Germany for support of the
construction and operation of the GEO600 detector. The authors also gratefully
acknowledge the support of the research by these agencies and by the Australian
Research Council, the Council of Scientific and Industrial Research of India,
the Istituto Nazionale di Fisica Nucleare of Italy, the Spanish Ministerio de
Educaci\'on y Ciencia, the Conselleria d'Economia Hisenda i Innovaci\'o of the
Govern de les Illes Balears, the Royal Society,
the Scottish Funding Council, the Scottish
Universities Physics Alliance, The National Aeronautics and Space
Administration, the Carnegie Trust, the Leverhulme Trust, the David and Lucile
Packard Foundation, the Research Corporation, and the Alfred P. Sloan
Foundation.

This document has been assigned LIGO Laboratory document number LIGO-\ligodoc.

\end{acknowledgments}

\appendix

\section{Data quality flags}\label{sec:dataquality}

Data quality flags are defined by the LIGO Detector Characterization group
by carefully processing information on the behavior of the instrument
prior to analyzing unshifted triggers. 
Some are defined online, as the data are acquired, while others are 
formulated offline. A wide range of DQFs have been defined. The relevance of
each available DQF has been evaluated and classified into categories which are
used differently in the analysis, which we now describe.

Category 1 DQFs are used to define the data set 
processed by the search algorithms.
They include out-of-science mode, the 30 seconds before loss of lock, 
periods when the data are corrupted and periods when test signals are injected 
into the detector. They also include short transients
that are loud enough to significantly distort the detector response and 
could affect the power spectral density used for normalization by the search
algorithm, such as dropouts in the calibration and photodiode saturations.

Category 2 flags are {\it{unconditional}} post-processing data cuts, used to 
define the ``full'' data set used to look for detection candidates.
The flags are associated with unambiguous malfunctioning with
 a proven correlation with loud transients in the GW channel, where we
understand the physical coupling mechanism. 
They typically only introduce a fraction of a percent of deadtime over the run. 
Examples include
saturations in the alignment control system, glitches in the power mains,
time-domain calibration anomalies, and large glitches in the
thermal compensation system.

Category 3 DQFs are applied to define the ``clean'' data 
set, used to set an upper limit in the absence of a detection candidate.
Any detection candidate found at a time marked with a category 3 DQF
would not be immediately rejected but would be considered cautiously,
with special attention to the effect of the flagged condition on
detection confidence.
DQF correlations with transients in the GW channels are established at the
single interferometer level. Examples include the 120~s prior to lock-loss,
noise in power mains, transient drops in the intensity of the
light stored in the arm cavities,
times when one Hanford instrument is unlocked and may negatively
affect the other instrument,
times with particularly poor sensitivity, and times associated with severe seismic activity, high wind speed,
or hurricanes. These flags introduce up to $\sim$10\% dead time. 

Category 4 flags are advisory only: 
We have no clear evidence of a correlation to
loud transients in the GW channel, but if we find a detection candidate at
these times, we need to exert caution. Examples are certain data validation issues 
and various local events marked in the electronic logs
by operators and science monitors.

\begin{figure}
\includegraphics[width=3in]{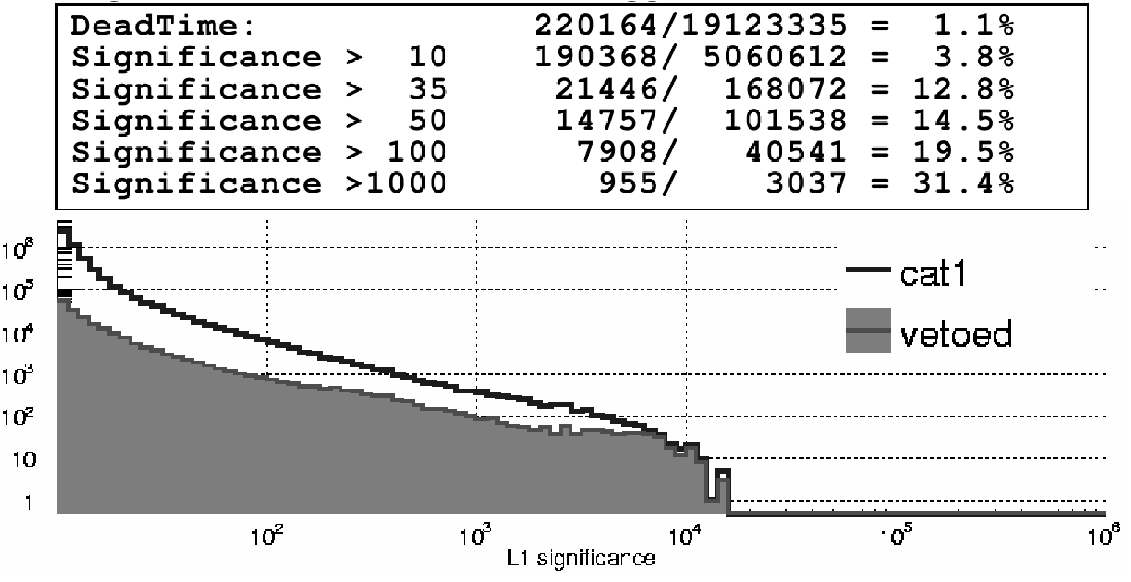}
\includegraphics[width=3in]{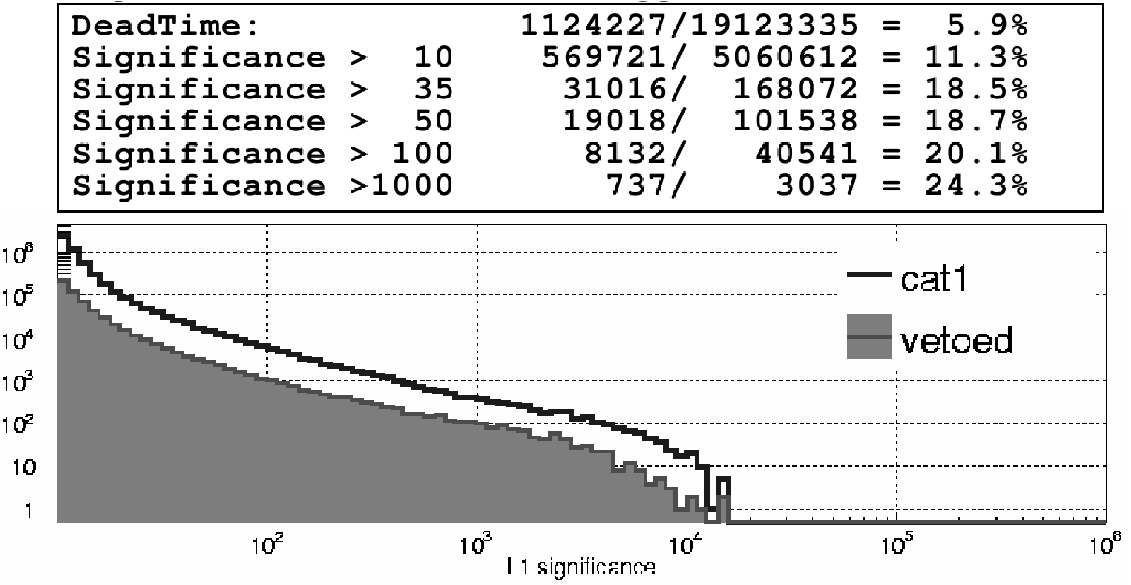}\\
\caption{The two examples in the figure show the fraction of single interferometer
(L1) KleineWelle triggers eliminated by category 2 (top) and
category 3 (bottom) DQFs, as a function of 
a threshold on the significance. 
The cumulative impact on the lifetime 
is less then  7 percent
(mostly from category 3 DQFs), and the cuts are most effective for the
loudest triggers.  For example, a significance of 1000 means that
if the detector noise were Gaussian, the noise would have a probability
$e^{-1000}$ of fluctuating to produce such a loud trigger.}
\label{figure:dqf}
\end{figure}

Figure~\ref{figure:dqf} shows the fraction of KleineWelle triggers that are eliminated by category 2 and 3 DQFs,
respectively, in the L1 interferometer, as a function of the 
significance of the energy excess identified by the trigger, which is evaluated
assuming stationary, random noise.
To ensure DQFs are independent of the presence of a true GW, 
we verified they are not triggered by hardware injections.

\section{Event-by-event vetoes}\label{sec:vetoes}

Event-by-event vetoes attempt to discard GW channel noise events
by using information from the many environmental and interferometric auxiliary
channels which measure non-GW degrees of freedom.  Good vetoes are found by
looking for situations in which a short ($\sim$ms) noise transient in an
auxiliary channel, identified by the KleineWelle (KW) algorithm,
often coincides within a short interval ($\sim$100 ms) with
noise transients in the GW channel. The work, then, is in
identifying useful auxiliary channels which are well correlated with noise
transients in the GW data, choosing the relevant veto parameters
to use, and finally establishing that the veto procedure will not systematically
throw out true GWs.
As for the data quality flags, vetoes are defined prior to generating triggers
from unshifted data.
The trigger properties used for veto studies are the KW signal
energy-weighted central time and the KW statistical significance.
The correlation between noise events in the GW channel and
an auxiliary channel is determined by a comparison of the coincidence
rate measured properly and coincidence rate formed when one of the time series
has been artificially time-shifted with respect to the other. Alternatively, we
can compare the number of coincidences with the number expected by chance,
assuming Poisson statistics.

As for the DQFs, category 2 vetoes are defined using
only a few subsets of related channels, showing the more obvious kinds of
mechanisms for disturbing the interferometers -- either vibrational or magnetic
coupling. Furthermore, for this S5 analysis we insist that multiple
(3 or more) channels from each
subset be excited in coincidence before declaring a category 2 veto, to ensure that a
genuine disturbance is being measured in each case. By contrast, the category 3
vetoes use a substantially larger list of channels. The aim of this latter
category of veto is to produce the optimum reduction of false events for a
chosen tolerable amount of livetime loss.

\subsubsection{Veto effectiveness metrics}
\label{vetostats}
{\em Veto efficiency} is defined for a given set of triggers as the fraction vetoed by our method.
We use a simple veto logic where an event is vetoed if its peak time falls within a veto window, and
define the {\em veto dead-time fraction} to be the fraction of livetime flagged
by all the veto windows. Assuming that real events are randomly distributed in
time, {\em dead-time fraction} represents the probability of vetoing a true
GW event by chance. We will refer to the flagged dead-time as
the {\em veto segments}.
A veto efficiency greater than the dead-time fraction indicates a
correlation between the triggers and veto segments.

Under either the assumption of randomly distributed triggers,
or randomly distributed dead-time, the number of events that fall within the
flagged dead-time is Poisson distributed with mean value equal to the
number of events times the fractional dead-time, or equivalently, the event rate
times the duration of veto segments. We
define the statistical significance of actually observing $N$ vetoed events as
$\mathrm{S}(N)=-\log{\left[\mathrm{P}_{\mathrm{Poiss}}(x \geq N)\right]}$.

We must also consider the {\em safety} of a veto condition:
auxiliary channels (besides the GW channel) could in principle be affected by a
GW, and a veto condition derived from such a channel could
systematically reject a genuine signal. Hardware signal injections imitating the
passage of GWs through our detectors, performed at several
pre-determined times during the run, have been used to establish under what
conditions each channel is safe to use as a veto. Non-detection of a hardware
injection by an auxiliary channel suggests the unconditional safety of this
channel as a veto in the search, assuming that a reasonably broad selection of
signal strengths and frequencies were injected. But even if hardware injections
are seen in the auxiliary channels, conditions can readily be derived under
which no triggers caused by the hardware injections are used as vetoes. This
involves imposing conditions on the strength of the triggers and/or on the ratio
of the signal strength seen in the auxiliary channel to that seen in the GW
channel.

Veto safety was quantified in terms of the probability of observing $\ge$N
coincidence events between the auxiliary channel and hardware injections vs. the
number of coincidences expected from time-shifts.

The observed concident rate is a random variable itself that 
fluctuates around the true coincident rate. 
In the veto analysis we use the 90\% confidence upper limit 
on the background coincidence rate which can be derived from the observed
 coincidence rate.
This procedure makes it easier to consider a veto safe than 
unsafe and the reason for this approach was to lean toward vetoing 
questionable events. A total of 20 time-shifts were
performed. The analysis looped over 7 different auxiliary channel thresholds and
calculated this probability, and a probability of less than 10\% caused a veto
channel at and below the given threshold to be judged unsafe.  A fixed 100 ms
window between the peak time of the injection and the peak time of the
KleineWelle trigger in the auxiliary channel was used.\\

All channels used for category 2 vetoes were found to be safe at any
threshold. Thresholds for category 3 veto channels were chosen so as
to ensure that the channel was safe at that threshold and above.

\subsubsection{Selection of veto conditions}

For the purpose of defining conservative vetoes appropriate for applying as
category 2 (before looking for GW detections), we studied
environmental channels. We found that these fall into groups of channels that
each veto a large number of the same events. Based on this observation, three
classes of environmental channels were adopted as vetoes. For LHO these classes were 
24 magnetometers and voltmeters with a KW threshold of 200 and time
window of 100 ms, and 32 accelerometers and seismometers with a threshold 
on the KW significance of
100 and a time window of 200 ms. For LLO these were 12 magnetometers and voltmeters
with a KW threshold of 200 and a time window of 100 ms. We used all of the
channels that should have been sensitive to similar effects across a site, with
the exception that channels known to have been malfunctioning during the time
period were removed from the list.

To ensure that our vetoes are based on true environmental disturbances, a further
step of {\em voting} was implemented. An event must be vetoed by three or more
channels in a particular veto group in order to be discarded from the detection
search. These conditions remove $\sim$0.1\% from the S5 livetime.

In the more aggressive category 3 vetoes, used for cleaning up the data for an
upper limit analysis, we draw from a large number of channels (about 60
interferometric channels per instrument, and 100 environmental channels per
site). This task is complicated by the desire to choose optimal veto thresholds
and windows, and the fact that the veto channels themselves can be highly
correlated with each other so that applying one veto channel changes the
incremental cost (in additional dead-time) and benefit (in additional veto
efficiency) of applying another. Applying all vetoes which perform well by
themselves often leads to an inefficient use of dead-time as dead-time continues
to accumulate while the same noise events are vetoed over and over.

For a particular set of GW channel noise events, we 
adopt a ``hierarchical'' approach to
choose the best subset of all possible veto conditions to use for a
target dead-time. This amounts to finding an ordering of veto conditions (veto
channel, threshold, and window) from best to worst such that the desired set of
veto conditions can be made by accumulating from the top veto conditions so
long as the dead-time does not exceed our limit, which is typically a few
percent.

We begin with an approximately ordered list based on the performance of each
veto condition (channel, window, and threshold) considered
separately. Incremental veto statistics are calculated for the entire list of
conditions using the available ordering. This means that for a given veto
condition, statistics are no longer calculated over the entire S5 livetime, but
only over the fraction of livetime that remains after all veto conditions
earlier in the list have been applied. The list is then re-sorted according to 
the incremental performance metric and the process is repeated until further
iterations yield a negligible change in ordering.

The ratio of {\em incremental veto efficiency} to {\em incremental dead-time} is
used as a performance metric to sort veto conditions. This ratio gives the
factor by which the rate of noise events inside the veto segments exceeds the
average rate. By adopting veto conditions with the largest incremental
efficiency/dead-time ratio, we maximize total efficiency for a target
dead-time. We also set a threshold of probability $P < 0.001$ on veto
significance (not to be confused with
the significance of the triggers themselves). This is particularly important for
low-number statistics when large efficiency/dead-time ratios can
occasionally result from a perfectly random process.

\begin{figure}
\includegraphics[width=3in]{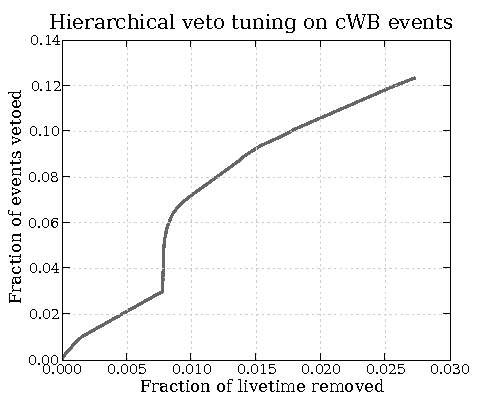}
\includegraphics[width=3in]{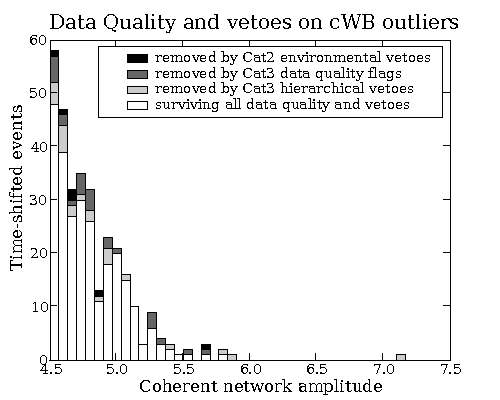}
\caption{Top: Accumulated veto efficiency versus dead-time as vetoes are applied
cumulatively down the veto list. The best vetoes are applied first, so we see a
general decrease in the effectiveness of vetoes at higher dead-time. Vetoes from
environmental channels are artificially prioritized over interferometric
channels, giving rise to the knee in the plot around 0.8\% deadtime where the
environmental vetoes are exhausted.  Bottom: Histogram of coherent network
amplitude, $\eta$, for coherent WaveBurst time-shifted (background)
events representing 100 S5 livetimes. The different shades
show events removed by data quality cuts and vetoes at various stages in the
analysis. }
\label{fig:cwbvetoeff}
\end{figure}

Vetoes were optimized over several different sets of GW channel
noise events including low-threshold H1H2L1 coherent WaveBurst time-shifted
events, H1H2 coherent WaveBurst playground events, as well as QPipeline and 
KleineWelle single-interferometer triggers.
For example, the effect of data quality flags and event-by-event
vetoes on the sample of coherent WaveBurst time-shifted events is
shown in Fig.~\ref{fig:cwbvetoeff}.
Our final list of veto segments to
exclude from the S5 analysis is generated from the union of these
individually-tuned lists.

\section{The BlockNormal Burst Search Algorithm}
\label{sec:blocknormal}

\subsection{Overview}

The BlockNormal analysis pipeline follows a similar logic to the
S4 burst analysis~\cite{S4burstAllSky} by looking
for bursts that are both coincident and correlated. The BlockNormal
pipeline uses a change-point analysis to identify coincident transient events of high significance
in each detector's data. The subsequent waveform correlation test is the same as that used in
the S4 analysis.

A unique feature of the BlockNormal analysis is that it can be run on uncalibrated time series data---neither
the change point analysis nor the correlation test are sensitive to the overall normalization of
the data. 

\subsection{Data conditioning}

The BlockNormal search operated on the frequency range 80 to 2048 Hz.
To
avoid potential issues with the additional processing and filtering used to create
calibrated data, and to be immune to corrections in the calibration procedure,
the analysis was run on the uncalibrated GW channel from the LIGO interferometers.

The data conditioning began with notch filters to suppress
out-of-band (below 80\,Hz or above 2048\,Hz) spectral features such as 
low-lying calibration lines, the strong 60 Hz power-line feature and violin mode 
harmonics just above 2048\,Hz.  The time-series data were then down-sampled 
to 4096 Hz to suppress high-frequency noise. The power-line harmonics in 
each band were removed using Kalman filters~\cite{Kalman60a, Finn:2000tt}.
The large amount of power at low frequencies in the uncalibrated GW channel 
was suppressed with a highpass filter designed with the Parks-McClellan algorithm.

Because the BlockNormal method is purely a time-domain statistic, the 
interferometer data must be divided into frequency bands to achieve a
degree of frequency resolution on the bursts.   For this analysis, 12 frequency bands 
approximately 150\,Hz in bandwidth spanned the range from 80\,Hz to 2048\,Hz 
(see Table~\ref{table:freqbands}).  
There are gaps between some bands to 
avoid the significant non-stationary noise from the violin modes of the mirror 
suspension wires.

\begin{table}
  \centering
  \begin{ruledtabular}
  \begin{tabular}{ccc}
    Lower Bound (Hz) & Upper Bound (Hz) & Bandwidth (Hz)\\ \hline
                80 & 192 & 112 \\
                192 & 320 & 128 \\
                362 & 518 & 156 \\
                518 & 674 & 156 \\
                710 & 864 & 154 \\      
                864 & 1018 & 154 \\
                1060 & 1212 & 152 \\ 
                1212 & 1364 & 152 \\
                1408 & 1558 & 150 \\ 
                1558 & 1708 & 150 \\
                1756 & 1902 & 146 \\ 
                1902 & 2048 & 146 \\ 
  \end{tabular}
  \end{ruledtabular}
        \caption{Frequency Bands for BlockNormal Analysis}
        \label{table:freqbands}
\end{table}

The division into the twelve frequency bands was done using a basebanding procedure. 
Any calibration lines within the band were removed by low-order regression filtering 
against the calibration line injection channel data.  A final whitening filter of modest order 
was applied in each band to satisfy the BlockNormal statistic's assumption of Gaussianity in the
background noise. The data conditioning procedures also had to minimize mixing noise characteristics 
between different time periods for the change-point analysis, and thus could 
not rely on predictive filtering.

\subsection{Change-point analysis}
The BlockNormal algorithm uses a Bayesian statistic termed $\rho_{2}$ to 
perform a change-point analysis using the noise characteristics of time-series 
data.  For an interval of $N$ time-series samples $x[k]$, this statistic measures
the statistical likelihood (at each sample $k$ within that interval) that the 
data prior to that point are more consistent with a different Gaussian-distributed 
(or normal) noise source than are the data following that point.  It is defined as 
\begin{eqnarray}
\label{eq:rho2exp}
\rho_{2,k} &=& K_{\rho} N \sqrt{\frac{\pi}{2}}
 \left[\frac{k^{-(k-1)/2} (N-k)^{-(N-k-1)/2}}{N^{-(N-1)/2}} \right]\\ \nonumber
&&\times \left[\frac{Y_{1,k}^{-k/2} Y_{k+1,N}^{-(N-k)/2}}{Y_{1,N}^{-N/2}}\right]
\left[\frac{\Gamma(k/2) \Gamma((N-k)/2)}{\Gamma(N/2)}\right]
\end{eqnarray}
where
\begin{eqnarray}
Y_{i,j} := \overline{x_{i,j}^2} - \overline{x_{i,j}}^2 \\
\overline{x_{i,j}} := \frac{1}{j-i+1} \sum_{l=i}^{j} x[l] \\
\overline{x_{i,j}^2} := \frac{1}{j-i+1} \sum_{l=i}^{j} x[l]^2 \,.
\end{eqnarray}
The quantity $K_{\rho}$ is a constant proportional to $\beta R / f_{s}$,
where $\beta$ is the prior probability, R the desired rate of blocks,
and $f_{s}$ the sample rate. 
In fact each interval is searched for all change-points where $\rho_{2,k}$ exceeds 
a threshold value $\rho_{E}$, where  $\rho_{E}$ is implemented as a number
times $K_{\rho}$.  
The sub-intervals between change-points are 
termed ``blocks''.  The statistical significance of each such block is based on its
``excess power'' $\xi^{*}$ defined as 
\begin{equation}
\xi^{*} = N \times (\mu^{2} + \nu) / (\mu_{0}^{2} + \nu_{0}) \sim \chi_{N}^{2}
\end{equation}
where the block has mean $\mu$ and variance $\nu$ against 
a background of mean $\mu_{0}$ and variance $\nu_{0}$. Events were selected  
by requiring the negative-log-likelihood of $\xi^{*}$ (termed $\Lambda_{E}$)
to exceed a threshold. Here
\begin{equation}
\Lambda_{E} =-ln({\mathop{\mathrm{Pr}}\nolimits}[\xi >\xi^{*}]) 
\end{equation}
where
\begin{equation}
{\mathop{\mathrm{Pr}}\nolimits}[\xi >\xi^{*}] = \gamma(N/2,\xi^{*}/2)/\gamma(N/2).
\end{equation}

The variance-weighted time centroid, $\tau^{(2)}$, of each event of $n$ samples of 
amplitude $x_{i}$ and time $t_{i}$ was calculated:
\begin{equation}
\tau^{(2)} = \frac{\sum_{i=1}^{n} t_{i}(x_{i}-\mu)^{2}}{\sum_{i=1}^{n}
  (x_{i}-\mu)^{2}}  
= \frac{\sum_{i=1}^{n} t_{i}(x_{i}-\mu)^{2}}{(n-1)\nu} \, .
\end{equation} 
The calibrated band-limited strain energy of each event was estimated using 
the frequency-averaged response $R(f)$ over that band:
\begin{equation}
E_{f} = R(f)(\mu^{2}n + \nu(n-1)) \, .
\end{equation}
The BlockNormal algorithm was applied separately to the data in each frequency band (Table~\ref{table:freqbands})
to select candidate GW burst events. The burst event generation was done on 
relatively long-duration epochs (up to 1200\,s) of continuous data to provide 
the best measure of the background noise characteristics.

Prior to the network coincidence step, events within each frequency band that are nearly 
adjacent were clustered into composite events.  Then, events between adjacent 
frequency bands whose time centroids were close were clustered into composite 
multiband events.  All events were then characterized by their frequency coverage. 
For composite events, the effective time centroid was the energy-weighted 
average of the time centroid of the constituent events.  The band-limited energy 
for composite events was simply the sum of the per-event energies.  The central 
frequency for events in a single band was estimated by the average frequency of 
that band. For multi-band events, the energy-weighted average of these central 
frequencies was used.

\subsection{Network coincidence}
The signals from actual GW bursts in the LIGO
interferometers should be separated in time by no more than the maximum transit
time ($10$\,ms) for GW between the Hanford and Livingston sites.
For the co-located interferometers at Hanford, there should be no time
separation. The separation observed in the reconstructed events is larger due to
limited time resolution, phase-delays in filtering, etc.  For a candidate
trigger, the time difference between candidate events in each pair of
interferometers, $|\tau_{i}^{(2)} - \tau_{j}^{(2)}|$, was required to fall
within a fixed coincidence window, $\Delta T_{ij}$, for that pair of
interferometers. This coincidence window had to be much broader  
than the transit time to account for limited time resolution and 
skewing of the time distributions from differential antenna response 
to $h_{+}$ and $h_{\times}$ waveforms.

The signals from actual GW bursts should also have similar strain amplitude (and
hence statistical significance) in each interferometer.  We derived a measure of
coincident significance from the excess power significance $\Lambda_{E}$ in each
candidate event in the trigger.  This measure must correct for the lower
significance for GW signals in the shorter H2 interferometer (as compared to the
H1 interferometer) as well as the fluctuation of the relative GW signal
strengths at the two LIGO sites due to modulation from the antenna factors.  The
chosen metric for coincident significance, termed ``combined power'' or $P_{\text{C}}$,
was defined as
\begin{equation}
P_{\text{C}} = (\Lambda_{E,\text{H1}}\Lambda_{E,\text{H2}}\Lambda_{E,\text{L1}})^{1/3} \,.
\end{equation}
This formulation was found to have the best performance in optimizing sensitivity to GW burst
signals as a function of the background trigger rate.

The coincidence procedure first identified events from each of the three detectors 
that had overlapping frequency coverage.  These events then had to have 
time centroids whose difference $\Delta T$ was less than $100$\,ms.  Such 
time-coincidence events were retained as GW burst triggers if their 
combined power $P_{\text{C}}$ was above a threshold of 22.

\subsection{Network correlation}
The signals from GW bursts in each interferometer result from the same parent
waveforms, and thus should have a large correlation sample-by-sample (after
correction for propagation delay).  The cross-correlation statistic $\Gamma$
reported by the CorrPower~\cite{Cadonati:2004ms} package is the maximum of the
average correlation confidence of pair-wise correlation tests. It is
positive-definite.  Larger values denote greater statistical certainty of
coherence. The CorrPower package was run on the list of candidate trigger times
produced in the coincidence step. It retrieved the full time-series data from
each interferometer around that time, calibrated the
data, and
calculated the $\Gamma$ cross-correlation statistic. For the three LIGO
interferometers, cuts were also made on the three pair-wise correlation tests.

Additional selection criteria took advantage of the special
relationship for GW signals from the co-located interferometers H1 and H2. One
was the signed correlation factor between the H1 and H2 interferometers from the
CorrPower processing, termed $R_0$.  For triggers from GW bursts, this
correlation factor should be positive.  For triggers from a background of random
coincidences, there should be an equal number of positive and negative
correlation factors.  Also, since the H1 and H2 interferometers receive
the same GW signal,
the ratio of $h_{\text{rss,H2}}$ to $h_{\text{rss,H1}}$ should be close to one
for a true GW burst.  In contrast, for triggers from a random background
this ratio will be centered around one-half. This arises because the H2
interferometer is approximately half as sensitive as H1, so signals of
the same statistical significance (near the threshold) will have only
one-half the amplitude in H2 as they do in H1.  To
simplify thresholding, the absolute value of the logarithm of the ratio was
calculated $R_{\text{H1H2}} = | \log_{10} ( h_{\text{rss,H1}} / h_{\text{rss,H2}} ) | $ for later
use.

\begin{figure}[htbp]
\begin{center}
\includegraphics[height=6cm]{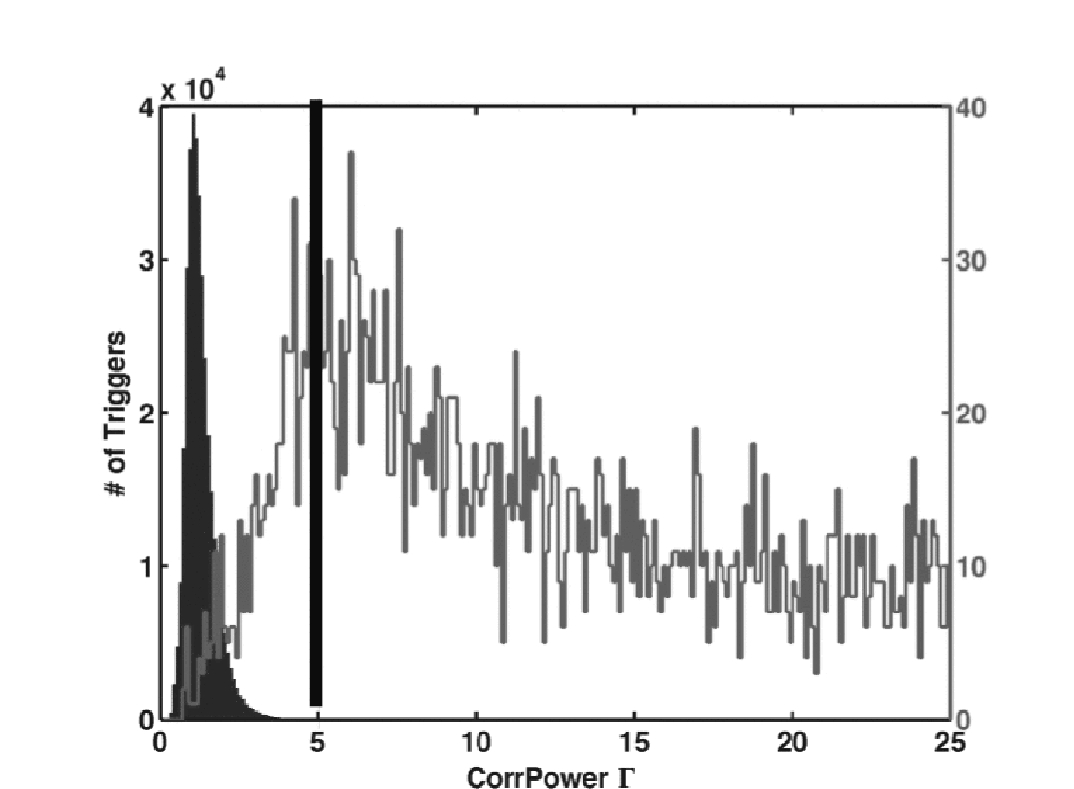} 
\caption{\small 
Distribution of background
and injection events with respect to the CorrPower $\Gamma$.
The narrow black histogram represents the background (noise) triggers while
the broader histogram represents the distribution of the injections.
These triggers were generated in the H1H2L1 network 
and contain frequencies below 200\,Hz.
The vertical line indicates the cut made on this quantity.} 
\label{fg:tuningBN}
\end{center}
\end{figure}

The choices of tuning parameters are described in Table~\ref{tab:tuningcpbn}.
Figure \ref{fg:tuningBN} illustrates an example of plots used to tune the figures
of merit for the H1H2L1 network.

\begin{table}[htbp]
  \caption{\label{tab:tuningcpbn} Cuts used by the BlockNormal-CorrPower
   pipeline in the first year of S5. The parameters are:
      combined power $P_{\text{C}}$,
      overall CorrPower $\Gamma$ value,
      CorrPower $\Gamma$ values for various detector pairs,
      H1-H2 correlation $R_0$,
      and estimated $h_{\text{rss}}$ values in H1 and H2.}

\begin{ruledtabular}
\begin{tabular}{l}
 \multicolumn{1}{c}{H1H2L1 Network} \\\hline
 $P_{\text{C}}>2$ \\
 $\Gamma>5.0$ for $f<200$\,Hz  \\
 $\Gamma>3.8$ for $f>200$\,Hz \\
 $\Gamma_{\text{H1H2}}>0.5$,
 $\Gamma_{\text{H1L1}}>0.3$,
 $\Gamma_{\text{H2L1}}>0.3$ \\
 $R_0>0$ \\
 $|\log_{10}(h_{\text{rss,H1}}/h_{\text{rss,H2}})|<0.4$ \\
\end{tabular}
\end{ruledtabular}

\end{table}

\section{The QPipeline burst search algorithm}
\label{sec:qpipeline}

\subsection{Overview}

QPipeline is an analysis pipeline for the detection of GW bursts in data from
interferometric gravitational wave detectors~\cite{chatterjiThesis}.  It is
based on the Q transform~\cite{multiresolution}, a multi-resolution
time-frequency transform that projects the data under test onto the space of
bisquare-windowed complex exponentials characterized by central time $\tau$,
central frequency $f_0$, and quality factor $Q$:

\begin{equation}
X(\tau,f_0,Q)  = \int_{-\infty}^{+\infty}
\tilde{x}(f) \tilde{w}(f,f_0,Q) e^{+i2\pi f \tau} df \, ,
\end{equation}
where the bisquare window $\tilde{w}(f,f_0,Q)$ is
\begin{equation}
  = \left\{
  \begin{array}{ll}
  \displaystyle A \left[1-\left(\frac{f Q}{f_{0} \sqrt{5.5}}\right)^2\right]^2
  & \text{for}~\displaystyle f < \frac{f_0\sqrt{5.5}}{Q} \\
  0 & \text{otherwise}
  \end{array}
  \right.   
\end{equation}
with
\begin{equation}
  A = \left(\frac{315}{128 \sqrt{5.5}}\frac{Q}{f_0}\right)^{1/2} \,.
\end{equation}

The bisquare window is a close approximation to a Gaussian window in frequency space;
the QPipeline is effectively a templated matched filter
search~\cite{wainstein62} for signals that are Gaussian enveloped sinusoids in
the whitened signal space.

\subsection{Data conditioning}

Before applying the Q transform, the data are first whitened by zero-phase
linear predictive filtering~\cite{makhoul,chatterjiThesis}.
In linear predictive whitening, the $n$th sample of a discrete data sequence is
assumed to be well modeled by a linear combination of the previous $M$ samples:
\begin{equation}
\hat{x}[n] = \sum_{m=1}^{M} c[m] x[n-m] \, .
\end{equation}
The resulting whitened data stream is the prediction error sequence $e[n] =
\hat{x}[n] - x[n]$ that remains after selecting the coefficients $c[m]$ to
minimize the error in the least-squares sense.

The prediction error length $M$ is taken to be equal to the length of the
longest basis function under test, which is approximately 1 second.  This
ensures that the data are uncorrelated on the time scales of the analysis.

In order to avoid introducing phase errors between detectors, a modified
zero-phase whitening filter is constructed by zero-padding the initial filter,
converting to the frequency domain, and discarding all phase information.

\subsection{Measurement basis}

The space of Gaussian enveloped complex exponentials is an over-complete basis of
waveforms, whose duration $\sigma_t$ and bandwidth $\sigma_f$ have the minimum
possible time-frequency uncertainty, $ \sigma_t \sigma_f = 1 / 4 \pi$, where $Q
= f_0 / \sqrt{2}\sigma_f$.  As a result, they provide the tightest possible constraints
on the time-frequency area of a signal, maximizing the measured signal to noise
ratio (SNR) and minimizing the probability that false triggers are coincident in
time and frequency between multiple detectors.

In practice, the Q transform is evaluated only for a finite number of basis
functions, which are more commonly referred to as templates or tiles.  These
templates are selected to cover a targeted region of signal space, and are
spaced such that the fractional signal energy loss $-\delta Z/Z$ due to the
mismatch $\delta \tau$, $\delta f_0$, and $\delta Q$ between an arbitrary basis
function and the nearest measurement template,
\begin{equation}
\frac{-\delta Z}{Z} \simeq
\frac{2 \pi^2 f_0^2}{Q^2} \, \delta \tau^2 +
\frac{1 + Q^2}{2 f_0^2} \, \delta f_0^2 +
\frac{1}{2 Q^2} \, \delta Q^2 -
\frac{1}{f_0 Q} \delta f_0 \, \delta Q,
\end{equation}
is no larger than $\sim\!\!20\%$.  This naturally leads to a tiling of the
signal space that is logarithmic in $Q$, logarithmic in frequency, and linear
in time.

For this search, the QPipeline was applied to search the space of sinusoidal
Gaussians with central frequency from 48\,Hz to 2048\,Hz, and with
$Q$ from $\sqrt{5.5}$ to $100/\sqrt{2}$.

\subsection{Trigger generation}

The statistical significance of Q transform projections are given by their
normalized energy $Z$, defined as the ratio of squared projection magnitude to
the mean squared projection magnitude of other templates with the same central
frequency and $Q$.  For the case of ideal white noise, $Z$ is exponentially
distributed and is related to the matched filter SNR quantity
$\rho$~\cite{wainstein62} by the relation
\begin{equation}
Z = |X|^2 / \langle |X|^2 \rangle_{\tau} = - \ln{\mathop{\textrm{Pr}}\nolimits}[Z^{\prime} > Z] = \rho^2 / 2 \,.
\label{Zdef}
\end{equation}

The Q transform is applied to the
whitened data and normalized energies are computed for each measurement
template as a function of time.  Templates with statistically significant signal
content are then identified by applying a threshold on the normalized energy.
Finally, since a single event may potentially produce multiple overlapping
triggers due to the overlap between measurement templates, only the most
significant of overlapping templates are reported as triggers.

Clustering of nearby triggers is not used in evaluating the significance of
events.  As a result, the detectability of GW burst signals depends on their
maximum projection onto the space of Gaussian enveloped sinusoids.

\subsection{Coherence}

For this search, the QPipeline took advantage of the co-located nature of
the two LIGO Hanford detectors to form two
linear combinations of the data streams from the two detectors.  This coherent
analysis makes use of correlations in the data to distinguish true GW signals from
instrumental glitches.

\subsubsection{Coherent signal stream}
The first combination is the coherent signal stream, $\text{H}+$, a frequency dependent weighted sum of the data from
the Hanford detectors which maximizes the effective SNR.  The weighting is inversely
proportional to the noise power spectral density, $S(f)$\,:

\begin{equation}
\tilde{x}_{\text{H}+}(f) = \left(\frac{1}{S_{\text{H1}}}+\frac{1}{S_{\text{H2}}}\right)^{-1}\left(\frac{\tilde{x}_{\text{H1}}(f)}{S_{\text{H1}}(f)} + \frac{\tilde{x}_{\text{H2}}(f)}{S_{\text{H2}}(f)}\right)
\label{CohStream}
\end{equation}

The resulting combination is treated as the output of a new hybrid, ``coherent''
detector.  Under the assumption that the power spectral density is approximately flat across
the window bandwidth, applying the Q transform to this data stream leads to a coherent energy value, $|X^{\text{coh}}_{\text{H}+}|^{2}$,
which takes the following form:
\begin{equation}
\begin{array}{rcl}
|X^{\text{coh}}_{\text{H}+}|^{2} & = & \left(\frac{1}{S_{\text{H1}}}+\frac{1}{S_{\text{H2}}}\right)^{-2} \times \\
  & & \left(\frac{ |X_{\text{H1}}|^{2} }{ S_{\text{H1}}^{2} } + \frac{ |X_{\text{H2}}|^{2} }{ S_{\text{H2}}^{2} }
+ \frac{X^{\ast}_{\text{H1}}X^{\vphantom{\ast}}_{\text{H2}} + X^{\vphantom{\ast}}_{\text{H1}}X^{\ast}_{\text{H2}}}{S_{\text{H1}} S_{\text{H2}}}\right)
\end{array}
\label{CohDef}
\end{equation}
where $X_{\text{H1}}$, $X_{\text{H2}}$, and $X^{\text{coh}}_{\text{H}+}$ are functions of $\tau$, $f_0$, and $Q$, and the asterisk denotes complex conjugation.
The last term represents the contribution of the cross-term, and is conceptually similar to a frequency domain representation of a cross-correlation of the H1 and H2 data streams.

The energy expected in the coherent data stream if there were no correlations in the data can be characterized by the ``incoherent'' terms
in Eq.~(\ref{CohDef}):
\begin{equation}
|X^{\text{inc}}_{\text{H}+}|^{2} = \left(\frac{1}{S_{\text{H1}}}+\frac{1}{S_{\text{H2}}}\right)^{-2}\left(\frac{ |X_{\text{H1}}|^{2} }{ S_{\text{H1}}^{2} } + \frac{ |X_{\text{H2}}|^{2} }{ S_{\text{H2}}^{2} }\right) \,.
\label{IncDef}
\end{equation}
The coherent and incoherent energies can then be normalized in the manner of Eq.~(\ref{Zdef}):
\begin{eqnarray}
Z^{\text{coh}}_{\text{H}+} &=& |X^{\text{coh}}_{\text{H}+}|^{2} / \langle |X^{\text{coh}}_{\text{H}+}|^{2} \rangle_{\tau}
\label{CohNorm}\\
Z^{\text{inc}}_{\text{H}+} &=& |X^{\text{inc}}_{\text{H}+}|^{2} / \langle |X^{\text{inc}}_{\text{H}+}|^2 \rangle_{\tau}
\end{eqnarray}
The correlation between the detectors can then be measured by the correlated energy, $Z^{\text{corr}}_{\text{H}+}$, given by
\begin{equation}
Z^{\text{corr}}_{\text{H}+} = Z^{\text{coh}}_{\text{H}+} - Z^{\text{inc}}_{\text{H}+} \simeq
 \frac{X^{\ast}_{\text{H1}}X^{\vphantom{\ast}}_{\text{H2}} + X^{\vphantom{\ast}}_{\text{H1}}X^{\ast}_{\text{H2}}}{S_{\text{H1}} + S_{\text{H2}}} \,.
\label{cor_energy}
\end{equation}

\subsubsection{Null stream}
The second combination is the difference between the calibrated data from the two detectors,
known as the null stream, and is defined as
\begin{equation}
\tilde{x}_{\text{H}-}(f) = \tilde{x}_{\text{H1}}(f) - \tilde{x}_{\text{H2}}(f).
\end{equation}
By subtracting the co-located streams, any true gravitational
wave signal should be canceled.  The resulting combination is treated as the output of a new hybrid ``$\text{H}-$''
detector, which shows significant energy content in the presence of instrumental glitches but does not respond to
gravitational waves.  Glitches are identified by thresholding on the
corresponding normalized ``null energy'', $Z^{\text{coh}}_{\text{H}-}$, calculated in an analogous manner to $Z^{\text{coh}}_{\text{H}+}$.

Signal tiles found to be in coincidence with significant null stream tiles are vetoed as instrumental glitches, and are
not considered as candidate events.  The threshold on $Z^{\text{coh}}_{\text{H}-}$ can be expressed as
\begin{equation}
Z^{\text{coh}}_{\text{H}-} > \alpha + \beta Z^{\text{inc}}_{\text{H}-}
\end{equation}
where $\alpha$ is chosen to limit the veto rate in Gaussian noise to $\sim1$
per 2048 tiles and $\beta$ is a parameter corresponding
to the allowed tolerance in calibration uncertainty. This is an energy factor, and corresponds to an
amplitude calibration uncertainty of approximately 22 percent.

We expect that highly energetic instrumental glitches could leak energy into
adjacent time-frequency bins, so the veto coincidence requirement between signal
and null streams is scaled to give more-significant null stream tiles more area
of veto influence in time-frequency space:
\begin{equation}
|\tau_{\text{H}-} - \tau_{\text{H}+}| < (\delta \tau_{\text{H}-}' + \delta \tau_{\text{H}+}) / 2 \,,
\end{equation}
\begin{equation}
|f_{0,\text{H}-} - f_{0,\text{H}+}| < (\delta f_{0,\text{H}-}' + \delta f_{0,\text{H}+}) / 2
\end{equation}
where $\tau$ and $f_0$ are the central time and frequency of
a tile, $\delta \tau$ and $\delta f$ are the duration and bandwidth of a tile, and
the inflated null stream tile duration and bandwidth are defined as:
\begin{equation}
\delta \tau_{\text{H}-}' = \max\left(1, 0.5  \sqrt{2  Z^{\text{coh}}_{\text{H}-}}\right) \times \delta \tau_{\text{H}-} \,,
\end{equation}
\begin{equation}
\delta f_{0,\text{H}-}' = \max\left(1, 0.5  \sqrt{2  Z^{\text{coh}}_{\text{H}-}}\right) \times \delta f_{0,\text{H}-}
\end{equation}

\subsection{Coincidence}

Coherent triggers from the two LIGO Hanford detectors were also tested for
time-frequency coincidence with triggers from the LIGO Livingston detector
using the following criteria, where $T$ is the speed of light travel time
of 10\,ms between the two LIGO sites:
\begin{equation}
|\tau_{\text{H}} - \tau_{\text{L}}| < \max(\delta \tau_{\text{H}}, \delta \tau_{\text{L}}) / 2 + T \,,
\end{equation}
\begin{equation}
|f_{0,\text{H}} - f_{0,\text{L}}| < \max(\delta f_{0,\text{H}}, \delta f_{0,\text{L}}) / 2 \,.
\end{equation}

Coincidence between the LIGO Hanford and Livingston sites is not a requirement
for detection, even if detectors at both sites are operational.  The final
trigger set is the union of triggers from the coherent H1H2 trigger set and the
coincident H1H2L1 trigger set.  The additional requirement of coincidence
permits a lower threshold, and therefore greater detection efficiency, for the
H1H2L1 data set.

\begin{figure}[htbp]
\begin{center}
 \includegraphics[height=6cm]{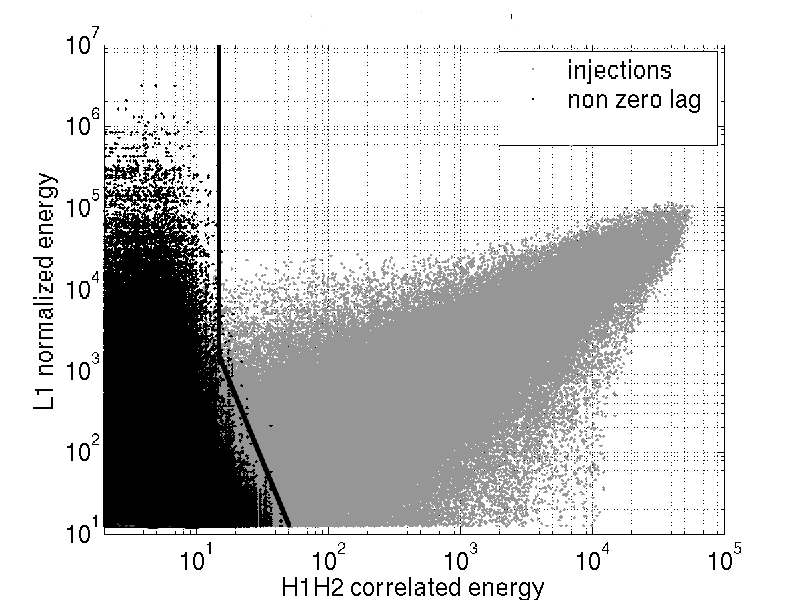}
\caption{\small
Scatter plot of the H1H2 correlated energy $Z^{\text{corr}}_{\text{H}+}$,
[defined in Eq.~(\ref{cor_energy})],
which measures the correlation of the strain at the two Hanford
interferometers, versus the L1
normalized energy [defined in Eq.~(\ref{Zdef})].
The distribution
of the background triggers is displayed in black while the distribution of simulated GW signals
in gray.
This example tuning plot is for triggers generated for the H1H2L1 network
and containing frequencies below 200\,Hz.
The cuts on these quantities are displayed on the plot as thick lines.
}
\label{fg:tuningQ}
\end{center}
\end{figure}

The choices of tuning parameters are described in Table~\ref{tab:tuningq}.
Figure \ref{fg:tuningQ} an example scatter plot used to tune the figures
of merit for the H1H2L1 network.

\begin{table}[htbp]
  \caption{\label{tab:tuningq} Cuts used by the QPipeline analysis in the
   first year of S5. The parameters are:
    H1/H2 coherent significance $Z^{\text{coh}}_{\text{H}+}$,
    H1/H2 correlated significance $Z^{\text{corr}}_{\text{H}+}$, and
    L1 normalized energy $Z_{\text{L1}}$. }

\begin{ruledtabular}
\begin{tabular}{@{\hfill}cll@{\hfill}c}
 \multicolumn{4}{c}{H1H2L1 Network} \\\hline
 & $Z^{\text{coh}}_{\text{H}+} > 20$  &  & \\
 & $\displaystyle Z^{\text{corr}}_{\text{H}+}>\max\left(15,50\sqrt[4]{\frac{12.5}{Z_{\text{L1}}}}\right)$ & for $f<200$\,Hz & \\
 & $\displaystyle Z^{\text{corr}}_{\text{H}+}>\max\left(5,30\sqrt{\frac{12.5}{Z_{\text{L1}}}}\right)$ & for $f>200$\,Hz & \\
 & $Z_{\text{L1}} > 12.5$ && \\ \hline\hline
 \multicolumn{4}{c}{H1H2 Network} \\\hline
 & $Z^{\text{coh}}_{\text{H}+} > 20$ &  & \\
 & $Z^{\text{corr}}_{\text{H}+}>50$ &for $f<200$\,Hz &\\
 & $Z^{\text{corr}}_{\text{H}+}>30$ &for $f>200$\,Hz &
\end{tabular}
\end{ruledtabular}

\end{table}

\section{The coherent WaveBurst search algorithm}
\label{sec:cWB}
\subsection{Overview}

Coherent WaveBurst (cWB) is an analysis pipeline for the detection and reconstruction
of GW burst signals from a network of detectors. The reconstructed
gravitational waveform ${\bf h}$ that best describes the response of the network
is used to compute the maximum likelihood ratio
of the putative GW signal, which forms the main detection statistic for the search.
In effect, cWB is equivalent to a matched filter search with a very large template
bank representing all possible time-domain signals with short duration.

The cWB pipeline is divided into three main stages: the generation of
coherent triggers, the reconstruction of the GW signal and the
computation of the maximum likelihood ratio, and a post-production
stage where additional detection cuts are applied.
By using weighted coherent combinations of the data streams, cWB is
not limited by the least sensitive detector in the network.
The waveform reconstruction allows various physical properties of the signal to
be estimated, including the sky location of the source. The coherent approach also
allows for other statistics to be constructed, such as the null stream and coherent energy,
to distinguish genuine GW signals from environmental and instrumental artifacts.

\subsection{Data conditioning and time-frequency decomposition}

The cWB analysis is performed in the wavelet domain. 
A discrete Meyer wavelet transformation is applied to the sampled detector output 
 to produce a discrete wavelet series  $a_k[i,j]$, where $i$ is the time index, 
$j$ is the scale index and $k$ is the detector index. An important property of 
Meyer wavelets is that they form an orthonormal basis that allow for the construction 
of wavelet filters with small spectral leakage~\cite{Klimenko:2008fu}. Wavelet series give 
a time-scale representation of data where each wavelet scale can be associated with a certain
frequency band of  the initial time series. Therefore a wavelet time-scale spectrum can be displayed as 
a time-frequency (TF) scalogram, where the scale is replaced with the central frequency 
$f$ of the band. The time series sampling rate $R$ and the scale number $j$ determine 
the time resolution $\Delta{t_j}(R)$ at this scale. The frequency resolution $\Delta{f_j}$ 
is defined as $1/(2\Delta{t_j})$ and determines the data bandwidth at the scale $j$.
The time-frequency resolution defines the tiling of the TF plane. The individual tiles 
(pixels) represent data samples in the wavelet domain. In the cWB pipeline a uniform 
tiling is used $(\Delta{f_j}(R)=R/2^n$, where $n$ is the wavelet decomposition depth), 
which is obtained with the Meyer packet transformation~\cite{wavelet}.
In this case the TF resolution is the same for all wavelet scales.  
For optimal localization of the GW energy in the TF plane, the cWB analysis is performed 
at six different frequency resolutions: 8, 16, 32, 64, 128 and 256 Hz. 

Before the coherent analysis is performed, two data conditioning algorithms are applied
to the data in the wavelet domain: a linear prediction error (LPE) filter and a wavelet 
estimator of the power spectral density $S_k[j]$.
LPE filters are used to remove ``predictable'' components                
from an input data series. In the cWB pipeline they are constructed individually for 
each wavelet layer and remove such components in the data as power line
harmonics and violin-mode lines. 
A more detailed description of the LPE filters can be found elsewhere~\cite{Klimenko:2005xv,Klimenko:2008fu}.
The wavelet estimator of the one-sided
power spectral density associated with each wavelet layer $j$ is
\begin{equation}
               S_k[j] = 2\frac{\sigma^2_k[j]}{R}
\end{equation}
where $\sigma^2_k[j]$ is the variance of the detector noise. In the analysis we assume
that the detector noise is Gaussian and quasi-stationary. The variance estimator may
vary with time and therefore it is calculated for each sample in the wavelet layer: 
$\sigma^2_k[i,j]$. The estimation of the noise variance is performed on data segments 
of length 60 seconds, with 40 seconds overlap. Linear interpolation is used 
between two measurements to obtain $\sigma^2_k[i,j]$.

\subsection{Coherent triggers}
\label{ctrig}

The first step in the analysis is to identify segments of data that may
contain a signal. The triggers are evaluated using the whitened data vector ${\bf{w}}[i,j]$
\begin{equation}
\label{eq:vec1}
{\bf{w}}[i,j](\theta,\phi) = \left( \frac{a_1[i,j,\tau_1(\theta,\phi)]}{\sigma_1[i,j]},..,
\frac{a_K[i,j,\tau_K(\theta,\phi)]}{\sigma_K[i,j]} \right) \;.
\end{equation}
The sampled detector amplitudes in the wavelet domain $a_k[i,j,,\tau_k]$ take into account the time delays
$\tau_k$ due to the time-of-flight between the detectors, which in turn depend on the source coordinates 
$\theta$ and $\phi$. Coherent triggers are generated for the entire network by maximizing the
norm $|{\bf{w}}[i,j]|$ over the entire sky for each time-frequency location $[i,j]$.
To do this, the sky is divided into square degree patches and the quantity $|{\bf{w}}|$ is calculated for each patch from 
the delayed detector amplitudes $a_k[i,j,\tau_k]$.
By selecting clusters of pixels with the $\max_{\theta,\phi}{|{\bf{w}}|}$ above some
threshold, one can identify coherent triggers in the time-frequency plane.
The data pixels $w_k[i,j]$ selected by this procedure are then used to reconstruct
the GW signal and compute the maximum likelihood statistic.

\subsection{Maximum likelihood ratio functional}
\label{CL}

For the case of Gaussian quasi-stationary noise, the likelihood that data ${\bf a}$ is purely instrumental
noise is proportional to $\exp\{- (a\vert a)/2\}$, while the likelihood that a GW signal ${\bf h}$ is present is proportional to $\exp\{- (a-h \vert a-h)/2\}$. The ratio of these
likelihoods can be used as a detection statistic. Here $(x\vert y)$ defines a noise
weighted inner product, which for $K$ detectors with uncorrelated noise can be written in the wavelet domain as
\begin{equation}
\label{eq:nwip}
(x\vert y) = \sum_{k=1}^K \sum_{i,j\in{\Omega_{TF}}} \frac{ x_k[i,j] y_k[i,j]}{\sigma^2_k[i,j]} \, .
\end{equation}
where time $i$ and frequency $j$ indices run over some time-frequency area $\Omega_{TF}$ selected for the analysis.
The coherent WaveBurst pipeline defines ${\cal L}$ as twice the (log) likelihood ratio, and treats it  as a functional in $h_{\text{det}}({\bf h})$
~\cite{Klimenko:2005xv}:
\begin{equation}
\label{eq:like}
{\cal{L}}[{\bf h}] = 2(a \vert h_{\text{det}})- (h_{\text{det}} \vert h_{\text{det}}) \, ,
\end{equation}
where $h^k_{\text{det}}[i,j]$ are the detector responses 
(Eq.~\ref{eq:hdet}).
The network sensitivity is characterized by the noise-scaled antenna pattern vectors 
${\bf{f_+}}$ and ${\bf{f_{\times}}}$:
\begin{equation}
\label{eq:vec2}
{\bf{f_{+(\times)}}}[i,j] = 
\left( \frac{F_{1,+(\times)}(\vec\Omega, \Psi)}
{\sigma_1[i,j]},..,\frac{F_{K,+(\times)}(\vec\Omega, \Psi)}{\sigma_K[i,j]} \right) \;.
\end{equation}
Since the detector responses $h^k_{\text{det}}$ are independent of 
rotation by an arbitrary polarization angle in the wave frame, 
it is convenient to perform calculations in the dominant polarization frame (DPF)~\cite{Klimenko:2005xv}.
In this frame the antenna pattern vectors ${\bf{f_+}}$ and ${\bf{f_{\times}}}$ are orthogonal to each other:
\begin{equation}
\label{eq:f1f20}
({\bf{f_{+}}}(\Psi_{DPF}) \cdot {\bf{f_{\times}}}(\Psi_{DPF})) = 0
\end{equation}
and  we refer to them as ${\bf{f_1}}$ and ${\bf{f_2}}$ respectively.
The corresponding solutions for the GW waveforms, $h_1$ and $h_2$, are found by variation of 
the likelihood functional (Eq.~(\ref{eq:like})) that can be written as the sum of two terms,
${\cal{L}}={\cal{L}}_1+{\cal{L}}_2$, where
\begin{eqnarray}
\label{eq:like1}
{\cal{L}}_1 =  \sum_{\Omega_{TF}} \left[ 2({\bf{w}}\cdot{\bf{f_1}}) h_1 - |{\bf{f_1}}|^2 h^2_1 \right]  \;, \\
\label{eq:like2}
{\cal{L}}_2 =  \sum_{\Omega_{TF}} \left[ 2({\bf{w}}\cdot{\bf{f_2}}) h_2 - |{\bf{f_2}}|^2 h^2_2 \right]  \;. 
\end{eqnarray}
The estimators of the GW waveforms for a particular sky location
are then the solutions of the equations $\delta {\cal{L}}_1 / \delta h_1 =0$ and $\delta {\cal{L}}_2 / \delta h_2 =0$:
\begin{eqnarray}
\label{eq:syst1}
 h_1 = {({\bf{w}}\cdot{\bf{f_1}})}/|{\bf{f_1}}|^2 \;, \\
\label{eq:syst2}
 h_2 = {({\bf{w}}\cdot{\bf{f_2}})}/|{\bf{f_2}}|^2 \;. 
\end{eqnarray}
Note, the norms $|{\bf{f_1}}|$ and $|{\bf{f_{2}}}|$ characterize the network 
sensitivity to the $h_1$ and $h_{2}$ polarizations respectively.
The maximum likelihood ratio statistic for sky location $(\theta,\phi)$ is calculated by 
substituting the solution for ${\bf h}$
into ${\cal L}[{\bf h}]$. The result can be written as
\begin{equation}
\label{eq:lMax}
L_{\mathrm{max}}(\theta,\phi) =\sum_{n,m=1}^{K} L_{mn} 
=\sum_{n,m=1}^{K} \sum_{\Omega_{TF}} w_n w_m P_{nm} \,,
\end{equation}
where the matrix $P$ is the projection constructed from the components of the 
unit vectors ${\bf{e_1}}$ and ${\bf{e_2}}$ along the directions of 
the ${\bf{f_1}}$ and ${\bf{f_2}}$ respectively:
\begin{equation}
P_{nm}=e_{1n}e_{1m}+e_{2{n}}e_{2{m}} \, .
\end{equation}
The kernel of the projection $P$ is the {\it signal} plane defined by these two vectors. 
The null space of the projection $P$ 
defines the reconstructed detector noise which is referred to as the null stream.

The projection matrix is invariant with respect to the rotation in the signal plane where
any two orthogonal unit vectors can be used for construction of the $P_{nm}$.
Therefore one can select vectors ${\bf{u}}$ and ${\bf{v}}$ such that
$({\bf{w}}\cdot{\bf{v}})=0$
and
\begin{equation}
\label{eq:u}
P_{nm}=u_{n}u_{m} \, .
\end{equation}
The unit vector ${\bf{u}}$ defines the vector 
\begin{equation}
\label{eq:xi}
{\bf{\xi}}=({\bf{w}} \cdot {\bf{u}}){\bf{u}}
\end{equation}
whose components are estimators of the noise-scaled detector responses 
$h^k_{\text{det}}[i,j]/\sigma_k[i,j]$.

\subsection{Regulators}
\label{REG}

In principle the likelihood approach outlined above can be used for the reconstruction of the GW waveforms and
calculation of the maximum likelihood statistic. 
In practice the formal solutions (\ref{eq:syst1}), (\ref{eq:syst2}) need to be regularized by constraints 
that account
for the way the network responds to a generic GW signal~\cite{Klimenko:2005xv}. For example, the network may
be insensitive to GW signals with a particular sky location or polarization, resulting in an ill-posed inversion problem.
These problems are addressed by using regulators and sky-dependent penalty factors.

A classical example of a singular inversion problem is a network of aligned detectors where 
the detector responses 
$h^k_{\text{det}}$  are identical. In this case the algorithm can be constrained to search for 
one unknown function rather than for the two GW polarizations
$h_1$ and $h_2$, which span a larger parameter space. Note that in this case 
$|{\bf{f_2}}|=0$, Eq.~(\ref{eq:syst2}) is ill-conditioned and 
the solution for the  $h_2$ waveform cannot be found.  
Regulators are important not only for aligned detectors, but also for networks 
of misaligned detectors, for example, the LIGO and Virgo network.
Depending on the source location, the network can be much less sensitive 
to the second GW component ($|{\bf{f_2}}|^2<<|{\bf{f_1}}|^2$) and the 
$h_2$ waveform may not be reconstructable from the noisy data.

In the coherent WaveBurst analysis we introduce a regulator by
modifying the norm of the ${\bf{f_2}}$ vector:
\begin{eqnarray}
\label{eq:reg}
|{\bf{f'_2}}|^2 = |{\bf{f_2}}|^2 + \delta \, , 
\end{eqnarray}
where $\delta$ is a tunable parameter. 
For example, if $\delta=\infty$, the second GW component is entirely suppressed 
and the regulator corresponds to the ``hard constraint'' described in Ref.~\cite{Klimenko:2005xv}.
In this case the unit vector ${\bf{u}}$ (see Eq.~(\ref{eq:u})) is pointing along the
${\bf{f_+}}$ direction.
In the cWB analysis the parameter $\delta$ is chosen to be
\begin{equation}
\label{eq:delta}
\delta = \left(0.01+\frac{2}{|w|^2}\right)\sum_k{\frac{1}{\sigma_k^2[i,j]}} \,.
\end{equation}
This regulator is more stringent for weak events which are generated
by the pipeline at much higher rate than the loud events.

The introduction of the regulator creates an obvious problem for the construction of 
the projection matrix.
Namely, the vector ${\bf{e'_2}}={\bf{f_2}}/|{\bf{f'_2}}|$ and the corresponding vector ${\bf{u'}}$ 
obtained by rotation of ${\bf{e_1}}$ and ${\bf{e'_2}}$ in the signal plane are not unit vectors 
if $\delta \neq 0$. To fix this problem we re-normalize the vector ${\bf{u'}}$ to unity
and use it for calculation of the maximum likelihood ratio and other coherent statistics.

\subsection{Coherent statistics}
\label{cuts}

When the detector noise is Gaussian and stationary, the maximum likelihood $L_{\mathrm{max}}$ 
is the only statistic required for detection and selection of the GW events. In this case
the false alarm and the false dismissal probabilities are controlled by the threshold 
on $L_{\mathrm{max}}$ which is an estimator of the total SNR detected by the network. 
However, the real data are contaminated with instrumental and 
environmental artifacts and additional selection cuts should be applied to separate 
them from genuine GW signals~\cite{Klimenko:2008fu}. 
In the coherent WaveBurst method these selection cuts are based on
coherent statistics constructed from the elements of the likelihood and the null matrices.
The diagonal terms of the matrix $L_{mn}$ describe the reconstructed normalized incoherent energy.
The sum of the off-diagonal terms is the coherent energy $E_{\mathrm{coh}}$ detected by the network.

The next step is to optimize the solution over the sky. Often, depending  on the network configuration,
the reconstruction of source coordinates is ambiguous. For example,
for two separated detectors the relative time delay that yields
maximum correlation between the data streams corresponds to
an annulus on the sky. In this case, an ``optimal'' source location is
selected, where the reconstructed detector responses are the most consistent with the output detector 
data streams. To properly account for the directional sensitivity of the network the optimization over
sky locations has to be more than a simple maximization of $L_{\mathrm{max}}(\theta,\phi)$.
In the cWB analysis the statistic that is maximized has the form
\begin{equation}
L_{sky}(\theta,\phi) = L_{\mathrm{max}}~P_f~\mathrm{cc},
\end{equation}
where $P_f$ is the penalty factor and $\mathrm{cc}$ is the network correlation coefficient.
$P_f$ and $\mathrm{cc}$ are defined below in terms of the matrix $L_{mn} = \sum_{\Omega_{TF}} w_n w_m P_{nm}$ and the
diagonal matrices $E_{nm} = E_n \delta_{nm}$ and $H_{nm}= H_n \delta_{nm}$ which describe the normalized
energy in the detectors, and the normalized reconstructed signal energy 
(see Eq.~(\ref{eq:xi})), with
\begin{equation}\label{eq:energies}
E_{k}=\sum_{\Omega_{TF}}{w^2_k}, \quad H_k = \sum_{\Omega_{TF}}{\xi^2_k} \, .
\end{equation}

Ideally, the reconstructed signal energy in each detector $H_k$ should not significantly exceed the energy 
$E_k$. This requirement can be enforced by the constraint
\begin{equation}
\label{eq:ed}
\Lambda_k = \sum_{\Omega_{TF}}{w_k \xi_k - \xi_k^2} = 0 
\end{equation}
for each detector in the network. These constraints can be applied during the signal reconstruction by way of
Lagrange multipliers in the variational analysis, however this greatly increases the computational
complexity of the algorithm. A simpler alternative is to introduce a penalty factor $P_f$ that penalizes sky locations violating
the constraint equation~(\ref{eq:ed}):
\begin{equation}
P_f = {\rm min}\left\{ 1, \sqrt{\frac{E_1}{H_1}},...,\sqrt{\frac{E_K}{H_K}} \right\} \; .
\end{equation}
In addition to serving as a penalty factor in the position reconstruction, the ratio of reconstructed
and detector energy were also used as a post-production cut. Events with $P_f < 0.6$ were discarded,
as were events with large values of the network energy disbalance 
\begin{equation}
\Lambda_{NET} = \sum_k{\frac{\vert \Lambda_{k} \vert}{E_{\rm{coh}}}} \, ,
\end{equation}
and the H1-H2 energy disbalance
\begin{equation}
\Lambda_{HH} = \frac{\vert \Lambda_{H1} - \Lambda_{H2} \vert}{E_{\rm{coh}}} \, .
\end{equation}
The latter cut was found to be particularly effective at rejecting correlated
glitches in the two Hanford interferometers.

The network correlation coefficient is also used to weight the overall likelihood for each sky location.
It is defined as
\begin{equation}\label{eq:ltf1}
   \mathrm{cc} = \frac{E_{\mathrm{coh}}}{N_{\mathrm{null}}+|E_{\mathrm{coh}}|}
\end{equation}
where $N_{\mathrm{null}}$ is the sum of all elements in the null matrix
\begin{equation}\label{eq:Nmatrix}
   N_{nm} = E_{nm} - L_{nm} \, ,
\end{equation}
which represents the normalized energy of the reconstructed noise. 
Usually for glitches little coherent energy is detected and
the reconstructed detector responses are inconsistent with the
detector output, which results
in a large value for the null energy. In addition to helping select the optimal sky location,
the correlation coefficients $\mathrm{cc}$ are used for a
signal consistency test based on the comparison of 
the null energy and the coherent energy. 

\begin{figure}[htbp]
\begin{center}
 \centering
 {\Large \raisebox{4.6cm}{$\mathbf{\eta}$}}
 \includegraphics[width=7.0cm, viewport= 30 0 520 550, clip=true]{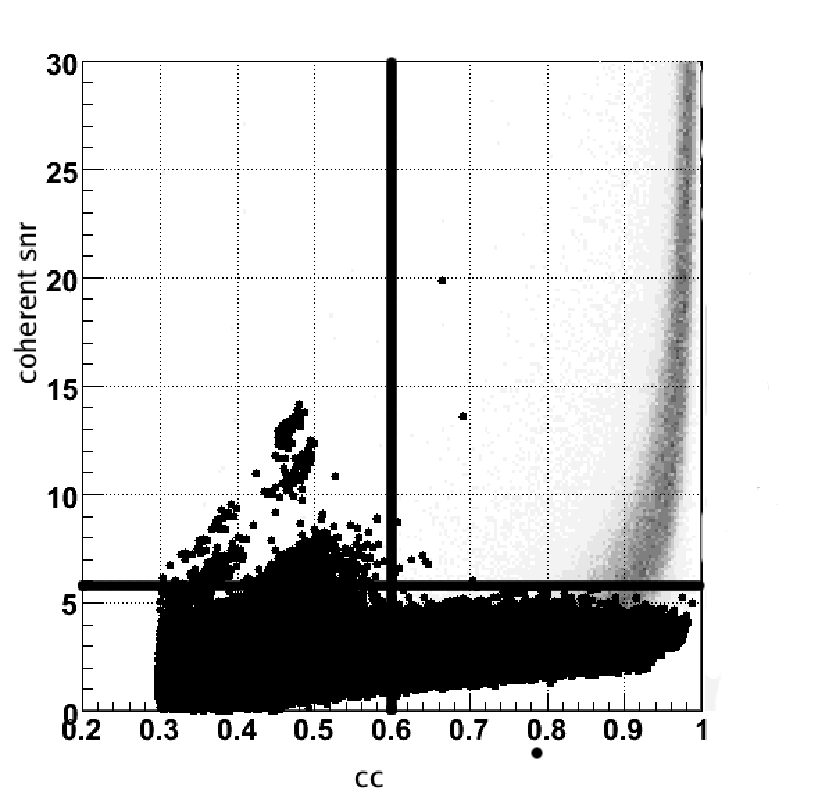}\\
\caption{\small Coherent network amplitude $\eta$ [defined in Eq.~(\ref{eq:rij2})]
versus network correlation coefficient $\mathrm{cc}$ [defined in (\ref{eq:ltf1})]
for cWB triggers below 200 Hz in the H1H2L1 network.
The black dots represent the noise triggers while the gray shadows represent 
the distribution of a set of simulated GWs injected into the data.
The horizontal and vertical bars represent the cuts on $\eta$ and $\mathrm{cc}$.} 
\label{fg:tuningCWB}
\end{center}
\end{figure}

The coherent terms of the likelihood matrix can be also used to calculate the 
correlation coefficients
\begin{equation}\label{eq:rij}
   r_{nm} = \frac{L_{nm}}{\sqrt{L_{nn}L_{mm}}}
\end{equation}
which represent Pearson's correlation coefficients in the case of aligned detectors.
We use the coefficients $r_{nm}$ to construct the reduced coherent energy
\begin{equation}\label{eq:rij1}
   e_{\mathrm{corr}} = \sum_{n\neq m}{L_{nm}\,|r_{nm}|} \,.
\end{equation}
Combined with the network correlation coefficient $\mathrm{cc}$ and
the number of detectors in the network, $K$, it yields a quantity
which we call the {\em coherent network amplitude},
\begin{equation}\label{eq:rij2}
   \eta = \sqrt{\frac{e_{\mathrm{corr}} \, \mathrm{cc}}{K}} \,.
\end{equation}
Figure~\ref{fg:tuningCWB} shows the $\eta$--$cc$ distribution of the background events 
(see Sec.~\ref{sec:tuning}) and simulated GW  events (see Sec.~\ref{sec:simulations}) 
for the L1H1H2 network. Loud background events due to detector glitches 
with low values of the network correlation coefficient are rejected by 
a threshold on $\mathrm{cc}$. Relatively weak background events are
rejected by a threshold on $\eta$.  
Table~\ref{tab:tuningcwb} describes the full set of tuning parameters for cWB.

\begin{table}[htbp]
  \caption{\label{tab:tuningcwb}Cuts used by the coherent WaveBurst pipeline
   in the first year of S5. The parameters are:
      network correlation coefficient $\mathrm{cc}$, 
      likelihood penalty factor $P_f$,
      coherent network amplitude $\eta$,
      H1-H2 energy disbalance $\Lambda_{HH}$, and
      network energy disbalance $\Lambda_\mathrm{NET}$.
      Time-dependent cuts are noted with UTC times.}

\begin{ruledtabular}
\begin{tabular}{l} 
 \multicolumn{1}{c}{H1H2L1 Network} \\ \hline
 $\mathrm{cc} > 0.6$ \, , \,\, $P_f > 0.6$ \\
 $\eta >5.7$ for $f$$<$$200$ Hz, up to Dec 12 2005 03:19:29\\
 \multicolumn{1}{r}{or after Oct 25 2006 09:34:17} \\
 $\eta >5.2$ for $f$$<$$200$ Hz, between Dec 12 2005 03:19:29\\
 \multicolumn{1}{r}{and Oct 25 2006 09:34:17} \\
 $\eta >4.25$ for $f$$>$$200$ Hz \\
 $\Lambda_\mathrm{HH} < 0.3 $ \, , \,\, $\Lambda_\mathrm{NET} < 0.35$ \\  \hline \hline
 \multicolumn{1}{c}{H1H2 Network} \\ \hline
 $\mathrm{cc} > 0.65$ \, , \,\, $P_f > 0.6$ \\
 $\eta >5.7$ for $f$$<$$200$ Hz \\
 $\eta >4.6$ for $f$$>$$200$ Hz, up to Jul 17 2006 11:50:37\\
 $\eta >4.25$ for $f$$>$$200$ Hz, after Jul 17 2006 11:50:37\\
 $\Lambda_\mathrm{HH} < 0.3 $ \, , \,\, $\Lambda_\mathrm{NET} < 0.35$ \\ \hline \hline
 \multicolumn{1}{c}{H1L1 Network} \\ \hline
 $\mathrm{cc} > 0.6$ \, , \,\, $P_f > 0.6$ \\
 $\eta >6.5$ for $f$$<$$200$ Hz, up to Oct 07 2006 08:58:06\\
 $\eta >9.0$ for $f$$<$$200$ Hz, after Oct 07 2006 08:58:06\\
 $\eta >5.0$ for $f$$>$$200$ Hz \\
 $\Lambda_\mathrm{NET} < 0.35$ \\  \hline \hline
 \multicolumn{1}{c}{H2L1 Network} \\ \hline
 $\mathrm{cc} > 0.6$ \, , \,\, $P_f > 0.6$ \\
 $\eta >6.5$ for $f$$<$$200$ Hz, up to Mar 28 2006 04:23:06\\
 \multicolumn{1}{r}{or after Oct 28 2006 11:54:46}\\
 $\eta >5.0$ for $f$$<$$200$ Hz, between Mar 28 2006 04:23:06\\
 \multicolumn{1}{r}{and Oct 28 2006 11:54:46} \\
 $\eta >5.0$ for $f$$>$$200$ Hz \\
 $\Lambda_\mathrm{NET} < 0.35$ \\
\end{tabular}
\end{ruledtabular}

\end{table}

\clearpage
\bibliographystyle{apsrev}

\end{document}